# Enhanced Simulation of the Indian Summer Monsoon Rainfall Using Regional Climate Modeling and Continuous Data Assimilation

Srinivas Desamsetti[1,2], Hari Prasad Dasari[2], Sabique Langodan[2], Yesubabu Viswanadhapalli[3], Raju Attada[4], Thang M. Luong[2], Omar Knio[2], Edriss S. Titi[5,6], and Ibrahim Hoteit[2,*]

[1]National Center for Medium-Range Weather Forecasting, Ministry of Earth Sciences, Noida, India
[2]Physical Sciences and Engineering Division, King Abdullah University of Science and Technology, Thuwal, Kingdom of Saudi Arabia
[3]National Atmospheric Research Laboratories, Gadanki, India
[4]Department of Earth and Environmental Sciences, Indian Institute of Science Education and Research, Mohali, India
[5]Department of Applied Mathematics and Theoretical Physics, University of Cambridge, Cambridge, United Kingdom
[6]Department of Mathematics, Texas A&M University, College Station, Texas, United States of America

**\*Corresponding author:** Prof. Ibrahim Hoteit,

King Abdullah University of Science and Technology (KAUST),

Physical Science and Engineering Division,

Thuwal 23955-6900, Saudi Arabia

E-mail: ibrahim.hoteit@kaust.edu.sa

**Abstract:**

This study assesses a Continuous Data Assimilation (CDA) dynamical-downscaling algorithm for enhancing the simulation of the Indian summer monsoon (ISM) system. CDA is a mathematically rigorous technique that has been recently introduced to constrain the large-scale features of high-resolution atmospheric models with coarse spatial scale data. It is similar to spectral nudging but does not require any spectral decomposition for scales separation. This is expected to be particularly relevant for ISM, which involves various interactions between large-scale circulations and regional physical processes.

Along with a control simulation, several downscaling simulations were conducted with the Weather Research and Forecasting (WRF) model configured over the Indian monsoon region at 10-km horizontal resolution using CDA, spectral (retaining different wavenumbers) and grid nudging for three contrasting ISM rainfall seasons: normal (2016), excess (2013), and drought (2009). The simulations are nested within the global NCEP Final Analysis data available at 1° x 1° horizontal resolution. The model outputs are evaluated against the India Meteorological Department (IMD) gridded precipitation and the fifth generation ECMWF atmospheric reanalysis (ERA-5).

Compared to grid and spectral nudging, the simulations using CDA produce enhanced ISM features over the Indian subcontinent including the low-level jet, tropical easterly jet, easterly wind shear, and rainfall distributions for all investigated ISM seasons. The major ISM processes, in particular the monsoon inversion over the Arabian Sea, tropospheric temperature gradients and moist static energy over central India, and zonal wind shear over the monsoon region, are all better simulated with CDA. Spectral nudging outputs are found to be sensitive to the choice of the wavenumber, requiring careful tuning to provide robust simulations of the ISM system. In contrast, control and grid nudging generally fail to well reproduce some of the main ISM features.

## 1. Introduction

The Indian summer monsoon rainfall (ISMR) occurs between June and September (JJAS) and contributes about 80% of the total annual Indian rainfall (e.g., Parthasarathy et al., 1994; Bollasina, 2014; Jain and Kumar 2012; Saha et al., 2014), playing a prominent role in India's agriculture and economy (Ghosh et al., 2016). Understanding and predicting ISMR is crucial for water resources management and planning crop strategies such as sowing, planting, irrigation, and developing best agricultural practices (Roy, 1990). Many models have been developed for simulating ISMR; however, these are yet to achieve highly robust predictive skills (e.g., Gadgil et al., 2005; Sahai et al., 2008; Raju et al., 2015a; Dash et al., 2015; Attada et al., 2019). Many studies reported that global climate models (GCM) have limited skills at representing ISMR main features such as the monsoon onset over Kerala (MOK), monsoon inversion, off-shore vortices over the west coast of India, monsoon trough, and the mesoscale features over the western ghats (e.g., Krishna Kumar et al., 2005; Kripalani et al., 2007; Dobler and Ahrens 2010; Raju et al., 2015a, b). Significant discrepancies in the GCMs simulated ISMR arise from their coarse resolution, mainly owing to inadequate representation of the orography and local mesoscale convective systems.

Regional Climate Models (RCMs) dynamically downscaling GCMs fields are the most popular tools to generate high resolution simulations of local atmospheric features over specific regions around the world (e.g., Giorgi and Mearns, 1999; Meehl et al., 2007; Rinke and Dethloff, 2000; Dasari et al., 2010, 2014; Attada et al., 2020). To constraint the RCM simulations to those of the underlying large-scale GCM features of interest, grid (Stauffer and Seaman, 1990) and spectral (Waldron et al., 1996; von Storch et al., 2000) nudging techniques have been applied. Grid nudging constrains the RCMs outputs to the global fields at every grid point of the domain. The balance of the RCMs, however, is better maintained by only constraining its large-scale features while allowing it to develop its own local variability; an approach known as spectral nudging (Miguez-Macho et al., 2005). The performance of spectral nudging strongly depends on the choice of the cut-off wave number, beyond which the scales freely evolve in the RCM (von Storch et al., 2000; Liu et al., 2012). Several studies assessed the characteristics of these two approaches for simulating different regional-scale features, including the ISM system (e.g., Miguez-Macho et al., 2005; Feser and von Storch, 2005; Winterfeldt and Weisse, 2009; Vincent and Hahmann, 2015; Desamsetti et al., 2019; Attada et al., 2021). Most of the ISM downscaling studies focused on understanding the rainfall characteristics and large-scale monsoon circulations (e.g., Park and Hong, 2004; Mukhopadhyay et al., 2010; Dasari et al., 2011; Srinivas et al., 2013; Devanand et al., 2018; Attada et al., 2020), but only few attempted to investigate the sensitivity of the simulated ISM to the downscaling techniques (Paul et al., 2016; Tang et al., 2017).

The Continuous Data Assimilation (CDA) approach was introduced to constrain dynamical models with available observations (Charney et al., 1969; Daley, 1991). It is generally implemented by adding a nudging term, function of model-data difference, to the underlying model predictions, similar to the nudging techniques (e.g., Henshaw et al., 2003; Olson and Titi, 2009; Hayden et al., 2011; Azouani et al., 2014). The nudging term in CDA is specifically designed to only constrain the large-scale variability of the model, so that the RCM is able to develop its own fine-scale features. The CDA framework constraining the asymptotic behavior of the spatial large-scale solution of the Navier-Stokes equations was shown to determine the asymptotic behavior of the full solution (Foias and Prodi, 1967; Foias and Temam, 1984; Jones and Titi, 1993; Cockburn et al., 1997; and references therein). More recently, Azouani et al. (2014) developed a CDA scheme in which nudging is performed on spatial interpolants of the model outputs towards the corresponding ones of the coarse data. This algorithm has been successfully tested in different settings, including the Navier-Stokes equations (e.g., Azouani et al. 2014, Farhat et al., 2016a; Mondaini and Titi, 2018; Biswas et al., 2019; Ibdah et al. 2020), Rayleigh-Bénard convection model (e.g., Farhat et al., 2018; Altaf et al., 2017), and planetary geostrophic ocean circulation model (Farhat et al., 2016b). It is worth mentioning that in certain settings it is sufficient to nudge only part of the state variables (Farhat et al., 2015, 2016a,b; Altaf et al., 2017). This algorithm has been further recently implemented for the first time with an atmospheric general circulation model (AGCM), the Weather Research and Forecasting (WRF) model, by Desamsetti et al. (2019), who demonstrated its advantage at better resolving both the large and small-scale features of the Arabian Peninsula climatic compared to the grid and spectral nudging techniques.

We investigate the performance of a WRF-CDA at simulating the ISM features for three contrasting ISMs reported by India Meteorological Department (IMD): normal (IMD, 2016), excess (IMD, 2013), and drought (IMD, 2009). IMD defines the normal, excess, and drought ISM seasons based on the percentage changes in the summer monsoon seasonal (June to September) rainfall over India compared to long-term precipitation. Along with the control simulation, we performed three experiments for each ISM season using WRF with one of the nudging techniques: grid, spectral, and CDA. We further performed four simulations using spectral nudging varying the cutoff wavenumber to investigate the sensitivity of the ISM features simulation to the choices of the wavenumber. The ISM simulations results are assessed in terms of resolving the monsoon onset over Kerala (MOK), rainfall distributions, circulation features, monsoon inversion over the Arabian Sea, moist static energy, and temperature gradient over central India.

The remainder of the study is organized as follows. Section 2 describes the model and nudging techniques, simulations setup, and datasets. The ISMR simulation results are presented and analyzed in Section 3. A summary of the main findings and a general discussion conclude the work in Section 4.

**2. Model and Datasets**

The simulations are performed using the Advanced WRF (WRF-ARW) model developed by the National Centers for Environmental Prediction (NCEP/NCAR) (Skamarock et al., 2019). WRF is configured here with 53 vertical levels and a single domain with 10-km horizontal resolution covering the ISM region (20 °E to 115 °E and 20 °S to 40 °N). The model topography is extracted from the 30s resolution Global Multi-resolution Terrain

Elevation Data (GMTED2010) developed by the United States Geological Survey and the National Geospatial-Intelligence Agency. The landuse, land cover categories are interpolated from the International Geosphere-Biosphere Programme 21-category data collected by the Moderate Resolution Imaging Spectroradiometer (MODIS IGBP) satellite. The initial conditions are obtained from the NCEP/NCAR Final Analysis (FNL) data available at 1° × 1°. The lower and lateral boundary conditions are also obtained from FNL data and updated every 6-hours. We investigated the three ISM seasons of 2016, 2013, and 2009 as representatives of the normal, excess, and drought ISM seasons, respectively. All simulations were performed over five months periods, starting from May to September. The first-month simulation is considered as a spin-up time and was not included in the analyses.

We performed seven WRF simulations for each of the three investigated seasons: 1) control run without nudging (WRF-CTL), 2) with grid nudging (WRF-GN), with spectral nudging (WRF-SN) retaining 3) three (WRF-SN33), 4) five (WRF-SN55), 5) seven (WRF-SN77), and 6) nine (WRF-SN99) wavenumbers in the x- and y-directions, and 7) with CDA (WRF-CDA). The WRF model physics are the same for all the simulations. The following parameterization schemes were selected for the simulations: Thompson microphysics scheme (Thompson et al., 2008), Kain-Fritch convective parametrization scheme (Kain, 2004), rapid radiative transfer scheme for both short wave and long wave radiations (Iacono et al., 2008), Mellor–Yamada surface layer scheme (Nakanish, 2001), and unified Noah land-surface model (Tewari et al., 2004). The nudging coefficient was set to 0.0003 s$^{-1}$ for all the nudged variables following Skamarock et al., (2019). Following previous works (e.g., Yesubabu et al., 2014; Raju et al., 2014; Gomez and Miguez, 2017; Desamsetti et al., 2019; Attada et al., 2021), WRF temperature, winds, along with geopotential height (for spectral and CDA nudging), and specific humidity (for grid nudging) were nudged towards the corresponding FNL data.

The downscaled rainfall and its variability during the three investigated monsoon seasons are evaluated against the IMD observed rainfall (Pai et al., 2014) available on a 0.25° × 0.25° grid. The simulated monsoon characteristics, including the mean circulation patterns at different pressure levels (950, 850, and 100 hPa), vertical wind shear (difference between 200 hPa and 850 hPa), the monsoon inversion (temperature difference between 850 and 950 hPa) over the Arabian Sea, tropospheric temperature gradients (temperature between 850 to 300 hPa), and moist static energy (MSE) for the studied normal, excess, and drought ISM seasons, are assessed and compared to those of the fifth-generation ECMWF reanalysis (ERA5) dataset (C3S, 2017; Hersbach et al., 2020).

## 3. Results

This section analyzes the monsoon characteristics as simulated by the WRF model with and without nudging in comparison with the observations.

### *3.1 Indian Summer Monsoon Rainfall Characteristics*

The rainfall characteristics, including the simulation of the monsoon onset over Kerala (MOK), rainfall distribution during the monsoon season (JJAS), and different phases of the monsoon such as progression (June), active (July and August), and withdrawal (September) are examined here.

### *3.1.1 Simulation of monsoon onset over Kerala (MOK)*

Simulating and understanding the monsoon onset over Kerala (MOK) is another important parameter as it signals the start of the rainy season over India, helping farmers to manage the pre-sowing activity and initiating the shift from a hot-dry season to a wet-rainy season. The climatological MOK of the ISM is 01 June with a standard deviation of 7 days (Krishnamurthy and Shukla, 2000). The IMD conventionally estimates the MOK based on rainfall measurements collected at meteorological stations deployed in Kerala (Ananthkrishnan and Soman, 1988; Soman and Kumar, 1993), however this method failed to detect the MOK of deadliest drought monsoon season (2002). Later, many studies attempted to estimate the MOK using various techniques (e.g., Webster and Yang, 1992; Goswami et al., 1999, 2006; Goswami and Gouda, 2007, Wang et al., 2001, 2009; Xavier et al., 2007; Amulya et al., 2019). Here, we investigate four objective methods (see Table 1) to estimate the MOK as suggested by Chevuturi et al. (2019): (i) Wang and Fan Index (WFI), based on the enhanced low-level circulation (Wang and Fan, 1999), (ii) Tropospheric Temperature Gradient Index (TTGI), based on the gradient of vertically averaged temperatures (Xavier et al. 2007), (iii) Webster and Yang Index (WYI), based on vertical wind shear (Webster and Yang, 1992), and (iv) Hydrological Onset and Withdrawal Index (HOWI), based on vertically integrated zonal moisture transport (Fasullo and Webster, 2003). Overall, the TTGI and WYI estimate the MOK based on large-scale features, whereas HOWI and WFI estimates are based on small-scale features.

The MOK estimated using the four different methods as resulting from the downscaling simulations of the investigated monsoon seasons is outlined in Table 2. IMD announced the MOK for the normal (2016), excess (2013), and drought (2009) years on 8 June, 1 June, and 23 May, respectively. The identified MOK by the four methods suggests that the CDA and spectral nudging experiments accurately simulate the MOK within ± 2 - 3 days of the observed MOK dates. The WRF-CTL experiment using Wang and Fan Index (Wang and Fan, 1999) simulated a delayed MOK by about 9 days (17 June) in the normal ISM, an early MOK by about 7 days (23 May) during the excess ISM, but agrees well with the IMD MOK date (23 May) during the drought monsoon season. Further, the increased west Asian surface pressure gradient between the Middle East and the West Equatorial Indian Ocean can potentially play a role in the occurrence of MOK through intensified low-level jet (Chakraborty and Agarwal, 2017). The distinct monsoon conditions exhibit (Figure 1) noticeable differences in the surface pressure gradients, which in turn modulate the onset dates of each monsoon season (Joseph et al., 1994; Chakraborty and Agrawal, 2017). Different downscaling techniques seem to have significant impact on reproducing the distinct patterns related to the different monsoon years. For instance, the reasonable representation of pressure gradients by CDA helps to achieve an accurate estimate of the onset dates compared to the standard nudging methods. Spectral nudging further showed important sensitivity to the choice of the wave numbers have in terms of the simulation of the pressure gradients and the corresponding onset dates. Spectral nudging simulations generally improved with higher number of waves and became comparable to CDA, when its number were retained. Overall, the dynamical downscaling simulations using the different nudging techniques successfully estimated the dates of MOK using the large-scale (TTGI and WYI) and small-scale (HOWI and WFI) indices, in agreement with the IMD. In contrast, WRF-CTL provided good estimate of MOK using the indices based on large-scale (TTGI and WYI), but not using those based on the small-scale features (HOWI and WFI), pointing to weaknesses at resolving ISM fine-scale features.

*3.1.2 Simulation of Rainfall distributions*

The mean rainfall averaged over a season (1 June to 30 September), progression (June), active (July and August), and withdrawal (September) phases during the investigated normal, excess, and drought monsoon seasons are examined for all WRF simulations (WRF-GN, WRF-CTL, WRF-CDA, WRF-SN99, WRF-SN77, WRF-SN55, WRF-SN33) and evaluated against the corresponding IMD rainfall distributions. The spatial distribution of IMD rainfall during the normal monsoon indicates that India's west coast and the northeastern parts, followed by central India, receive substantial rainfall of more than 10 mm/day (Figure 2, first row). The southern peninsular India (rain shadow region), north and northwest India receive relatively less rainfall of about 5 mm/day. WRF-CTL exhibits a significant dry bias over southern India and relatively less wet bias over the monsoon trough regions. WRF-CTL overestimates the rainfall during the normal monsoon year over central India, with a northward shift, and fails to capture the spatial variability of rainfall over Peninsular India. WRF-GN fails to reproduce these features, with significant wet bias both in terms of distribution and intensity over entire India, except for Jammu and Kashmir. These localized rainfall features are better simulated by WRF-CDA, followed by WRF-SN (SN99 followed by SN77, SN55, and SN33). WRF-CDA and WRF-SN99 runs reproduce the gross rainfall distributions, but show a dry bias over central India extending to northwest India, Tibet region, and Jammu and Kashmir regions. The two simulations further overestimated the rainfall over the northern parts of Madhya Pradesh and the western parts of Uttar Pradesh. Similar results were obtained for the simulation of the mean monthly rainfall distributions for the progression (i.e., June), peak (i.e., July and August), and withdrawal (i.e., September) phases of the monsoon (Supplementary Figure S1). Note that all the experiments failed to capture June rainfall, but with a relatively better performance from WRF-CDA and WRF-SN99.

WRF-CTL underestimates the rainfall during the excess monsoon year over Peninsular India, whereas it overestimates it over northwest India, Uttar Pradesh, Bihar. WRF-GN significantly overestimates the rainfall all over India, excluding Telangana (Figure 2, second row). WRF-CDA and WRF-SN overestimate the rainfall distribution all over India, except Gujarat, Western Ghats, Madhya Pradesh. Overall, WRF-CDA best reproduces the rainfall distribution compared to all other runs, relatively exhibiting less positive-bias over the dry regions such as the northwestern part of India and Rayalaseema. Although WRF-SN99, WRF-SN77, WRF-SN55, and WRF-SN33 are able to capture the spatial distribution of rainfall over most parts of India, these runs produced significantly higher rainfall over east-central India, the windward side of Eastern Ghats, and northeast India. Similar results are also obtained during the progression, active, and withdrawal phases of the monsoon (Supplementary Figure S2), with the best results again from WRF-CDA. WRF-CDA, followed by WRF-SN99, which further exhibited a weaker wet bias during the progression of the monsoon (June).

WRF-CTL well simulates (Figure 2, bottom row) the rainfall spatial distribution during the drought monsoon season (2009) over India, except over the high rainfall regions of the west coast of India. It is also able to reproduce the IMD distributions during the progression (June) and withdrawal (September) phases, but fails during the active phase (Supplementary Figure S3). WRF-GN fails to represent the rainfall distribution associated with the drought conditions, in all phases, as revealed by the overestimation of rainfall in comparison to all other runs. WRF-CDA is in good agreement with the IMD rainfall (Figure 2, third row), except over the foothills of the Himalayas and coastal regions of the eastern parts of India where it shows wet biases. WRF-SN99, WRF-SN77, WRF-SN55 and WRF-SN33 also produce similar spatial patterns as those of IMD, albeit for wet biases over West Bengal and northern parts of Odisha. The rainfall distributions associated with the drought conditions during the progression, active, and withdrawal phases (Supplementary Figure S3) over India are best reproduced by WRF-CDA, followed by WRF-SN99, WRF-SN77, WRF-SN55 and WRF-SN33, with the exception of the foothills of the Himalayas region.

We have further computed the differences between the area-averaged daily rainfall over central India (74.5°E to 86.5°E and 16.5°N to 26.5°N) as resulting from all downscaling runs and the IMD observations (Figure 3), and presented the correlations in Table 3. The time series of the rainfall differences of WRF-CTL show larger errors with IMD, simulating significantly more rainfall during some days and lower rainfall during others (Figure 3). The differences of WRF-GN with IMD show higher positive values, indicating that WRF-GN simulates more rainfall in all three ISM seasons. WRF-CDA seems to have the least bias and high correlation values (Table 3) with the IMD observations. The time-series of the rainfall differences between WRF-SN and IMD follow the patterns of those of WRF-CDA but with relatively higher biases and lower correlation values. The better rainfall simulations with the nudging methods suggest their relevance to better maintain the balance between the large-scale and small-scale features associated with ISMs. The rainfall analysis shows that WRF-CDA consistently produces more accurate rainfall over central India, the monsoon core region, compared to all other runs.

Overall, the performance of WRF-SN77 and WRF-SN55 in terms of MOK and rainfall distributions are somewhere between those of WRF-SN99 and WRF-SN33. Hereafter, we restrict our analysis to WRF-SN99 and WRF-SN33 to examine WRF-SN results of the ISMs features.

*3.2 Salient Features of the ISM*

*3.2.1 Monsoon Circulation*

The monsoon low-level jet (MLLJ), a southwesterly wind-flow confined to a narrow region over the western Arabian Sea, is a prime indicator of the ISM circulation (Joseph and Raman, 1966; Findlater, 1969). The MLLJ originates from cross-equatorial flow and acts as an important driver for transporting moisture towards the Indian region, controlling the monsoon inversion layers over the west central Arabian Sea (Dwivedi et al., 2016; 2021). The MLLJ further feeds moisture to the monsoon depressions formed over the Bay of Bengal (BoB) (Nagar et al., 2009; Walker et al., 2015) and modulate the rainfall over India (Roxy et al., 2017; Viswanadhapalli et al., 2020).

The spatial distribution of monsoon low-level (at 950 hPa) winds from the WRF simulations are presented along with ERA5 during the investigated normal, excess, and drought seasons (Figure 4 and Supplementary Figure S4). The WRF large-scale circulation features of ISM are similar to those of ERA5 at low levels during the normal and excess monsoon years, albeit to sharp reductions in the intensity of both cross-equatorial flow and MLLJ and an increase in the secondary branch of monsoon winds over the BoB. ERA5 also exhibits a noticeable variation in MLLJ strength and its north-south and east-west extensions during the normal, excess, and drought ISM seasons. The drought (excess) season is associated with weak (strong) MLLJ over the Arabian Sea, with smaller (higher) east-west extension and weak (strong) winds over the BoB compared to the normal season. The differences in the west Asian surface pressure gradients between Middle East and West Equatorial Indian Ocean during three different ISMs indicates larger (smaller) pressure gradients favor stronger (weaker) MLLJ during the excess (drought) ISM (Chakraborty and Agrawal, 2017). In all seasons, WRF-CTL and WRF-GN are able to simulate the majority of low-level wind features prevailing during ISM; however, the intensity of MLLJ and the southwesterly winds over BoB are significantly weaker than ERA5. All these circulation features are well reproduced in the WRF downscaling simulations, WRF-CDA, WRF-SN99, and WRF-SN33, and are in better agreement with ERA5 compared to WRF-GN and WRF-CTL. The east-west extension of the MLLJ in ERA5 is relatively higher in all monsoon seasons than in the WRF simulations. Similar results (not shown) are also obtained for the winds at 850 hPa.

The zonal winds and geopotential height at 850 hPa (averaged over monsoon core region, 50-60E) averaged between the monsoon period presented in Figure 5. The analysis from ERA5 suggest that there is a gradual increase of zonal wind starting from southern hemisphere and crossed the equator and attained its maxima between 5N-10N and thereafter start decreasing. The geopotential height is almost constant between the regions 5S to 5N, its start decreasing after 5N. All three ISM seasons show more or less similar features, however, with different magnitudes. Higher (lower) zonal winds noticeable between 5N-10N during excess (drought) ISM seasons. These features are consistently well represented by WRF-CDA and WRF-SN99 followed by WRF-SN33 than other simulations.

The mean circulation during the JJAS season at 100 hPa from ERA5 exhibits an anti-cyclonic circulation over the Tibetan Plateau and strong upper tropospheric easterlies, known as the tropical easterly jet stream (TEJ), over the Indian subcontinent (Figure 6 and Supplementary Figure S5). The strength of the TEJ is directly proportional to the strength of MLLJ (Tanaka, 1982) and was suggested to be an important factor in determining the monsoon rainfall over India (e.g., Chen and Van Loon, 1987; Chen and Yen, 1991; Pattanaik and Satyan, 2000). Consequently, the strength of the TEJ and their east-west extension are significantly higher during the excess monsoon seasons. Apart from reducing the TEJ intensity during the drought season, the location of its core also shifts from the west coast of India to the western BoB. WRF-CTL shows a significantly stronger and extended core region of TEJ compared to WRF with nudging during the excess monsoon season. The upper tropospheric large-scale mean circulation features in terms of the intensity of TEJ and location of its core are best simulated by WRF-CDA, followed by WRF-SN99, WRF-SN33, and WRF-GN, during the three different seasons. Overall, the upper-tropospheric circulation features such as Tibetan anticyclone, TEJ, and northward displacement of the subtropical westerly jet (SWJ) are well reproduced by all WRF runs and are in good agreement with those of ERA5.

*3.2.2 Horizontal Wind Shear*

The seasonal wind shear, generated by the low-level westerly and upper-level easterly winds, is a prominent monsoon feature of ISM (e.g., Xavier et al. 2007; Attada et al. 2018). It further contributes to the formation of the monsoon depression/low pressure systems over the BoB during the ISM and associated rainfall over central India (Naidu et al., 2011; Viswanadhapalli et al., 2020). The mean seasonal variations in vertical zonal wind shear as resulting from the different WRF downscaling experiments were analyzed and compared against those of ERA5 for all three studied ISMs (Figure 7 and Supplementary Figure S6). ERA5 shows a strong easterly shear over the north Indian Ocean during the normal and excess monsoon seasons compared to the drought season. Furthermore, ERA5 indicates a weaker zonal wind shear over a smaller region of the BoB (5°N–22°N and 80°E–100°E) during the drought season, and a wider strong easterly zonal wind shear region over the BoB during the normal and excess monsoon seasons. The strengthening of the zonal wind shear over the BoB decreases the formation of the number of depressions and cyclonic systems over the region and reduces rainfall over India, favoring dry spells or drought conditions (e.g., Naidu et al., 2011; Viswanadhapalli et al., 2020). The differences of the three different monsoon seasons as seen in ERA5 in terms of zonal wind shear are relatively well reproduced by WRF-CTL, WRF-GN, but are better represented in WRF-CDA followed by WRF-SN99 and WRF-SN33 (Figure 7). These deviations in vertical shear of zonal wind over the Indian Ocean induces significant changes in the atmospheric dynamics, which can eventually modulate the convective activities over the ISM region (Reale et al., 2017; Attada et al., 2018).

### *3.2.3 Monsoon Inversion over the Arabian Sea*

The depth and east-west extensions of the monsoon inversion (MI) over the western Arabian Sea is another distinct ISM feature that regulates the moisture transport towards the Indian landmass (Colon, 1964; Ramage, 1966). This also influences the exchange of moisture and energy between the ocean surfaces (Narayanan and Rao, 1981). The difference in temperature between the pressure levels of 850 and 950 hPa [$\Delta T = T_{850\ hPa} -$ $T_{950 hPa}$)] is used here to identify the MI (Dwivedi et al., 2016; 2021) for all WRF downscaling experiments during the three studied monsoon seasons and is compared with that of ERA5 (Figure 8). Muraleedharan et al. (2013) revealed that higher (lower) MI values over the western Arabian Sea are associated with the active (break) phases of the ISM. Dwivedi et al. (2016; 2021) reported that the strengthening of the MI is associated with a weak or a drought monsoon season. The analysis of MI as resulting from ERA5 suggests that the seasonal mean of MI is stronger over the west-central Arabian Sea during the drought season (Figure 8). WRF-CTL and WRF-GN show deviations of these features from ERA5, whereas similar features are reproduced by WRF-CDA, WRF-SN99, and WRF-SN33, albeit to a slight positive bias. The spatial extent from ERA5 indicates that the MI is confined to a very narrow region. The MI is weak during the excess monsoon season and the location its peak shifts toward the northwestern Arabian Sea during the normal monsoon season. WRF-CDA, WRF-SN99, and WRF-SN33 appreciably capture the MI spatial variability during the three different seasons, with a slight overestimation. We also analyzed the variations in the MI for each month (Supplementary Figure S7-S9) and noticed that the MI during the drought year (Supplementary Figure S9) is stronger in July and August compared to the other two monsoon seasons (Supplementary Figures S7 and S8). This suggests that the changes of MI during these two months are important for assessing the drought conditions over India. An eastward shift of the MI during July and August is further noticed towards the eastern Arabian Sea during the normal and excess ISM seasons (Supplementary Figures S7 and S8) in ERA5, WRF-CDA, WRF-SN99, and WRF-SN33. This suggests that the eastward extensions of MI favors more moisture transport from the Arabian Sea towards the Indian landmass.

### *3.2.4 Tropospheric Temperature Gradient (TTG) and Moist Static Energy (MSE)*

The tropospheric temperature gradients (TTG) between 850 to 300 hPa over the Indian Subcontinent during the monsoon season play an important role in determining the strength of the monsoon circulation (Parthasarathy et al., 1990; Singh and Chattopadhyay, 1998). The north-south temperature gradient manifests the position of the easterly shear, the dynamical feedback to the MLLJ, and the formation of the tropical easterly jet (Krishnamurti et al., 1976; Chen and van Loon, 1987; Raju et al., 2015a; Attada et al., 2018). The strong north-south tropospheric temperature gradient favors an increase in the thermal wind with height, leading to the monsoon circulation. The analysis of ERA5 indicates that the north-south tropospheric temperature gradients is significantly lower during the drought season compared to the normal and excess seasons (not shown). WRF-CTL and WRF-GN fail to reproduce the spatial patterns of the north-south tropospheric temperature gradients, in contrast with WRF-CDA, WRF-SN99, and to some extent with WRF-SN33, which reproduce noticeably well these features.

Figure 9 (left panel) shows the temporal evolution of TTG (averaged between 700 hPa and 200 hPa) gradient averaged over a region between 40°E–100°E; 5°N–35°N and 40°E–100°E; 15°S–5°N from ERA5 and the WRF simulations. ERA5 shows higher TTG during the peak monsoon months compared to the onset and withdrawal phases of the monsoon. The release of latent heat due to moisture condensation over the ISM region drives thermal forcing in the middle troposphere and strong TTG (e.g., Xavier et al., 2007, Attada et al., 2018; Attada et al., 2021). The change in latent heat due to moisture condensation modulates the intensity of the monsoon. The temporal evolution of TTG from ERA5 for all the monsoon seasons (i.e., normal, excess, and drought) shows a gradual increase in TTG during the onset phase, roughly maintaining same magnitude during progression, and a gradual decrease during the withdrawal phase. WRF-CTL overestimates the TTG relative to ERA5 as compared to WRF-GN. TTG with WRF-CDA and WRF-SN99, while it is slightly overestimated in WRF-SN33. This analysis confirms that the large-scale thermal contrast associated with the ISM is well represented in WRF-CDA and WRF-SN99.

Moist static energy (MSE) is the amount of latent energy due to water vapor present in the air parcel over central India and plays an important factor in the ISM variability. Raman and Rao (1981) and Krishnamurti et al. (2010) indicated that horizontal temperature advection modulates the MSE through to the intrusion of cold air. The MSE analysis suggests that relatively dry advection during the drought monsoon season leads to extended monsoon break conditions (Figure 9, right panel). A gradual increase during the onset phase, mostly constant during progression, and a gradual decrease during the withdrawal phases in MSE are noticeable during the normal and excess monsoon seasons. WRF-CTL and WRF-GN fail to reproduce the patterns of TTG and MSE compared to ERA5 for all three studied monsoon seasons, in contrast with WRF-CDA and WRF-SN99 which well reproduce those features as seen in ERA5.

*3.3 Quantitative Analysis*

We further analyzed the vertical distribution of the Root-Mean-Square-Error (RMSE) for winds, temperature, vertical velocity and relative vorticity over the monsoon core region as resulting from ERA5 and different WRF simulations (Figure 10). The RMSE between ERA5 and WRF-CDA and WRF-SN show relatively lower values than WRF-GN and WRF-CTL for the three studied ISM seasons. WRF-CTL has the highest RMSE over the entire troposphere for all dynamic and thermodynamic variables and all seasons, followed by WRF-GN. The RMSE of WRF-CDA is much lower for all variables in all three ISMs. WRF-CDA has better skills in the middle-to-upper (between 600-300 hPa) troposphere where the moist convective feedbacks are dominant. Similarly, in the case of temperature, RMSE of WRF-CTL is higher at the surface (more than 2K), whereas WRF-CDA has the least error (about 1K) followed by WRF-SN99 and WRF-SN33. WRF-CTL shows higher errors in vertical motions over the mid-troposphere with the least errors resulting again from WRF-CDA and WRF-SN99. This indicates that WRF-CDA represents well the monsoon-induced tropospheric heating and associated large scale ascents. Specifically, at lower levels, where the MLLJ prevails, the RMSE values resulting from WRF-CDA are significantly lower than those of all other WRF runs. These results clearly suggest that WRF-CDA, followed by WRF-SN99 and to some extent WRF-SN33, better resolve the vertical profiles compared to the other WRF runs for the investigated ISM seasons.

Overall, the conducted qualitative and quantitative analyses demonstrate that the newly implemented WRF-CDA methodology notably enhances the WRF model skill compared to the grid and spectral nudging methods in terms of simulating the deficit, normal, and excess ISM systems.

## 4. Summary and Conclusions

Current Indian Summer Monsoon (ISM) models face several challenges to properly describe the interactions between the large-scale and small-scale features, such as topography, land-sea contrast, mesoscale convective systems, monsoon depressions, and land-use distribution. This study assessed the performance of Continuous Data Assimilation (CDA) on the simulation of the Indian Summer Monsoon (ISM) system features with the Weather Research and Forecasting (WRF) model. With CDA, WRF (WRF-CDA) is able to produce enhanced downscaled information from the global NCEP/NCAR Final reanalysis (FNL). CDA achieves this by constraining the large spatial scales features through a nudging term, defined as the differences between interpolants of the model outputs and those of the data, which allows the child high resolution model to develop its own internal physics.

WRF-CDA simulations were compared against those of the control (WRF-CTL) and two standard downscaling methods: grid (WRF-GN) and spectral (WRF-SN) nudging in terms of simulating the features of a normal (2016), excess (2013), and drought (2009) ISM rainfall season. WRF simulations were conducted on a 10-km horizontal resolution grid and were nested within FNL. The India Meteorological Department (IMD) gridded precipitation fields and the fifth-generation ECMWF reanalysis (ERA5) datasets were used to evaluate the WRF simulations. The simulations were further examined in terms of the dynamical and thermodynamical monsoon features such as the onset over Kerala (MOK), the spatial characteristics of rainfall, the circulation features, the zonal wind shear over monsoon region, the monsoon low level jet (MLLJ), the monsoon inversion (MI) over Arabian Sea, and the tropospheric temperature gradients (TTG) and moist static energy (MSE).

Our results suggest that WRF-CDA best reproduced the ISM rainfall features compared to IMD observations during the normal, excess, and drought conditions over the Indian Subcontinent. WRF-CDA, followed by WRF-SN, well simulated the MOK and rainfall distributions. The spectral nudging experiments, constraining the first 9 wavenumbers in the zonal and meridional directions (WRF-SN99) produced better results compared to those resulting from WRF-SN constraining 7, 5, and 3 wavenumbers (WRF-SN77, WRF-SN55, and WRF-SN33). The simulated ISM rainfall features resulting from WRF-SN99 are overall in good agreement with WRF-CDA. WRF-GN and WRF-CTL mostly failed to simulate the normal and excess ISM rainfall features, even though the drought conditions are better resolved by WRF-CTL. The low-level cross-equatorial flow, MLLJ over Somalia coast, MI over the Arabian Sea, zonal wind shear, TTG over India, and MSE over central India are all best reproduced by WRF-CDA compared to those of WRF-CTL, WRF-GN and WRF-SN. The RSME values between the vertical profiles of winds, temperature, vertical velocity and relative vorticity over the monsoon core region from ERA5 and different WRF simulations are significantly lower in WRF-CDA, followed by WRF-SN99.

Our analysis clearly suggests that CDA provides enhanced WRF performance skills at simulating the ISM systems by maintaining a better balance between the large- and small-scale features.

**Acknowledgments:**

This publication is based upon work supported by the King Abdullah University of Science and Technology (KAUST) Office of Sponsored Research (OSR) under Award No. OSR-CRG2020-433, the Virtual Red Sea Initiative Award No. REP/1/3268-01-01, and the Saudi ARAMCO Marine Environmental Research Center at KAUST (SAMERK). The research made use of the Supercomputing Laboratory resources at KAUST.

Table 1: Details of the monsoon onset indices adopted in the study.

| Index & Reference | Methodology | Onset |
|---|---|---|
| Tropospheric Temperature Gradient Index (TTGI) Xavier et al. (2007) | TTGI = TN - TS  T = Vertically averaged temperature (600 - 200 hPa)  TN = T (40°E - 100°E, 5°N - 35°N)  TS = T (40°E -100°E, 15°S - 5°N) | Onset defined when TTGI time-series for each year becomes positive |
| Webster and Yang Index (WYI) Webster and Yang (1992) | WYI = U850 - U200  U = Zonal Wind at 850 or 200 hPa  U = U(40°E - 110°E, 0° -20°N) | Onset defined when seven-day running average of WYI timeseries crosses threshold value (mean ERA-Interim value on climatological onset, 30May) |
| Hydrological Onset and Withdrawal Index (HOWI) Fasullo and Webster (2003) | HOWI = nVIMT(1000 - 300 hPa)  nVIMT = Vertically integrated zonal moisture transport (300-1000hPa) normalized over the annual cycle  nVIMT = nVIMT(45°E - 80°E, 5°N - 20°N) | Onset defined when HOWI time-series for each year becomes positive |
| Wang and Fan Index (WFI) Wang and Fan (1999) | WFI = US – UN  U = Zonal Wind at 850 hPa  US = U(40°E - 80°E, 5°N - 15°N)  UN = U(70°E - 90°E, 20°N - 30°N) | Onset defined when seven-day running average of WFI timeseries becomes positive |

Table 2: The monsoon onset dates computed using different methods for all experiments along with the IMD reported onset dates.

| NORMAL - 2016 (IMD onset date 8-June) | | | | |
|---|---|---|---|---|
|  | TTGI | WYI | HOWI | WFI |
| WRF-CDA | 8-June | 8-June | 7-June | 7-June |
| WRF-SN99 | 8-June | 8-June | 7-June | 7-June |
| WRF-SN33 | 8-June | 8-June | 7-June | 7-June |
| WRF-GN | 7-June | 7-June | 6-June | 8-June |
| WRF-CTL | 10-June | 7-June | 7-June | 17-June |
| EXCESS - 2013 (IMD onset date 1-June) | | | | |
|  | TTGI | WYI | HOWI | WFI |
| WRF-CDA | 2-June | 1-June | 31-May | 28-May |
| WRF-SN99 | 2-June | 1-June | 31-May | 29-May |
| WRF-SN33 | 2-June | 1-June | 31-May | 30-May |
| WRF-GN | 2-June | 1-June | 30-May | 28-May |
| WRF-CTL | 31-May | 31-May | 30-May | 23-May |
| DROUGHT - 2009 (IMD onset date 23-May) | | | | |
|  | TTGI | WYI | HOWI | WFI |
| WRF-CDA | 25-May | 22-May | 23-May | 20-May |
| WRF-SN99 | 26-May | 22-May | 23-May | 20-May |
| WRF-SN33 | 25-May | 22-May | 23-May | 20-May |
| WRF-GN | 25-May | 21-May | 23-May | 21-May |
| WRF-CTL | 21-May | 21-May | 21-May | 21-May |

Table 3. The correlation coefficients between the model and observations. The rainfall is averaged over the monsoon core regions (74.5°E to 86.5°E and 16.5°N to 26.5°N).

|  | 2016 | | | | | 2013 | | | | | 2009 | | | | |
| --- | --- | --- | --- | --- | --- | --- | --- | --- | --- | --- | --- | --- | --- | --- | --- |
|  | JJAS | JUN | JUL | AUG | SEP | JJAS | JUN | JUL | AUG | SEP | JJAS | JUN | JUL | AUG | SEP |
| WRF-CTL | 0.47 | 0.76 | 0.58 | 0.13 | -0.09 | 0.41 | 0.07 | 0.14 | 0.42 | 0.46 | 0.38 | 0.09 | -0.21 | -0.05 | 0.28 |
| WRF-GN | 0.87 | 0.86 | 0.78 | 0.74 | 0.67 | 0.88 | 0.91 | 0.79 | 0.88 | 0.85 | 0.86 | 0.90 | 0.78 | 0.87 | 0.92 |
| WRF-CDA | 0.73 | 0.51 | 0.81 | 0.65 | 0.65 | 0.71 | 0.66 | 0.63 | 0.72 | 0.49 | 0.77 | 0.58 | 0.81 | 0.72 | 0.88 |
| WRF-SN99 | 0.70 | 0.38 | 0.81 | 0.67 | 0.69 | 0.69 | 0.65 | 0.64 | 0.67 | 0.40 | 0.75 | 0.43 | 0.81 | 0.77 | 0.86 |
| WRF-SN33 | 0.65 | 0.57 | 0.76 | 0.50 | 0.51 | 0.69 | 0.68 | 0.61 | 0.67 | 0.39 | 0.71 | 0.62 | 0.68 | 0.76 | 0.67 |

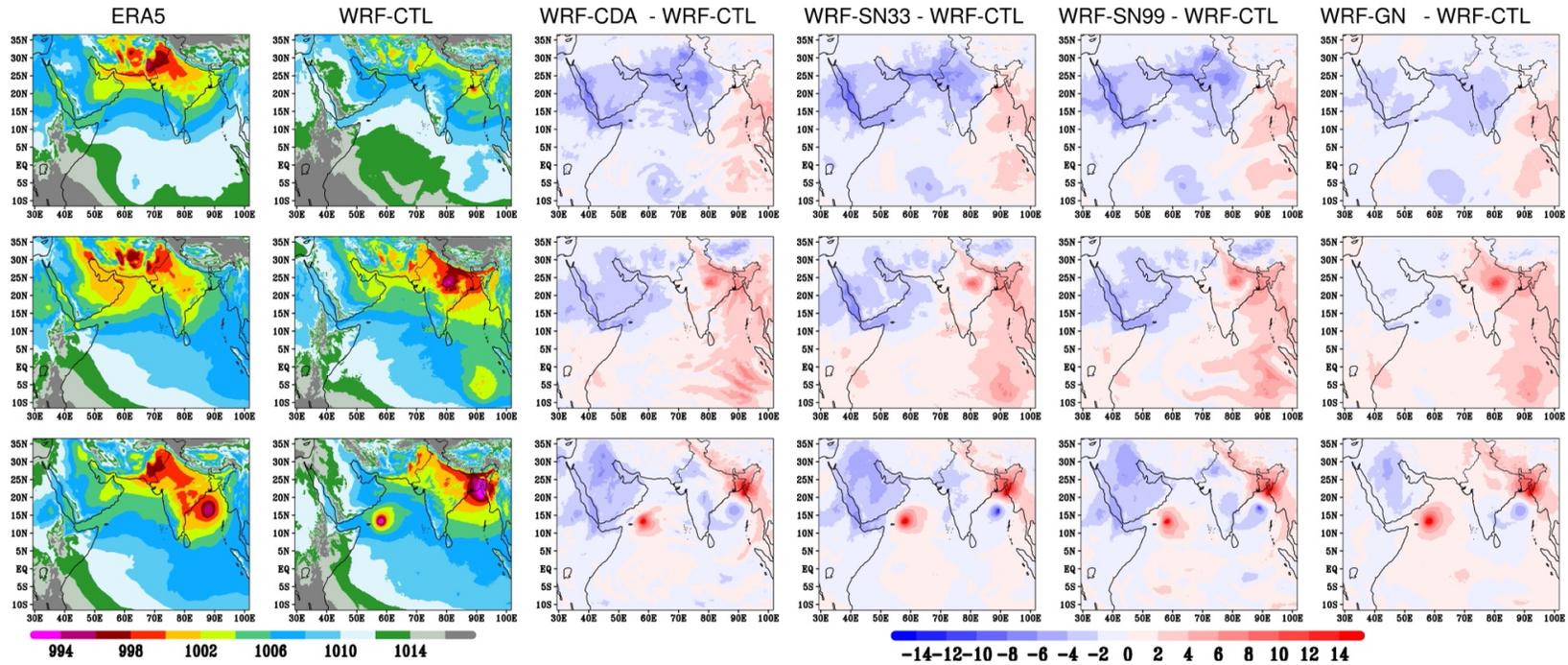

Figure 1. The mean seas level pressure (hPa) of ERA5 (First column) and WRF-CTL (Second column), and the differences in the mean sea level pressure between the WRF-CDA and WRF-CTL (Third column), WRF-SN33 and WRF-CTL (Fourth column), WRF-S99 and WRF-CTL (Fifth column), and WRF-GN and WRF-CTL (Sixth column) during normal (2016), excess (2013), and drought (2009) Indian Summer Monsoon seasons. First row is for normal, second row is for excess, and third row is for drought Indian Summer Monsoon seasons.

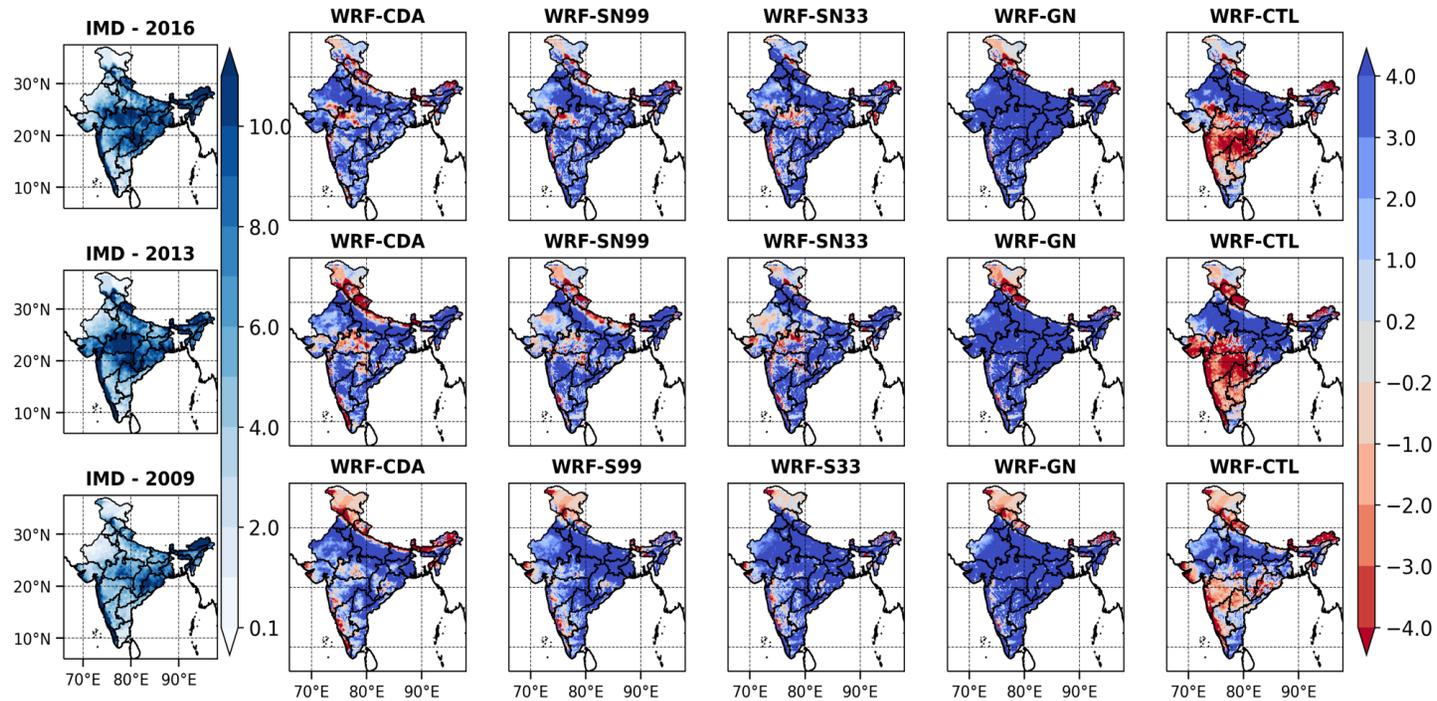

Figure 2. The India Meteorological observed mean seasonal rainfall (mm/day, First column) during normal (2016), excess (2013), and drought (2009) Indian Summer Monsoon seasons. Second, third, fourth, fifth, and sixth, seventh, and eighth columns are the seasonal mean rainfall differences between the model (WRF-CDA, WRF-SN99, WRF-SN77, WRF-SN55, WRF-SN33, WRF-GN, and WRF-CTL, respectively) and IMD observations for all three seasons. First row is for normal, second row is for excess, and third row is for drought Indian Summer Monsoon seasons.

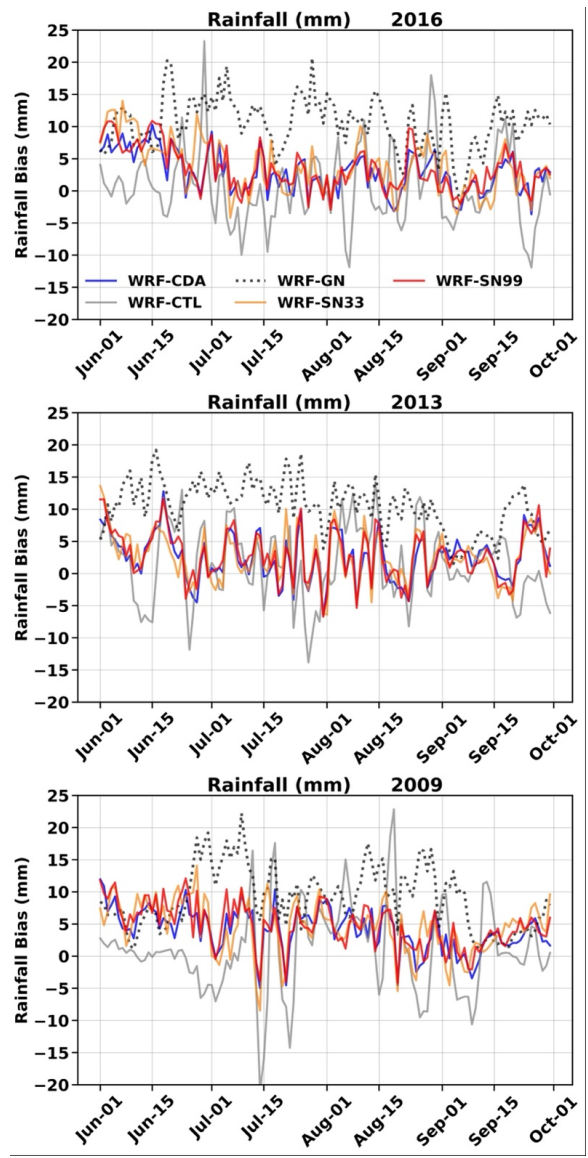

Figure 3. Time-series of the rainfall differences (Model-IMD) during normal (2016), excess (2013), and drought (2009) Indian Summer Monsoon seasons. The rainfall is averaged over the monsoon core region (74.5°E to 86.5°E and 16.5°N to 26.5°N).

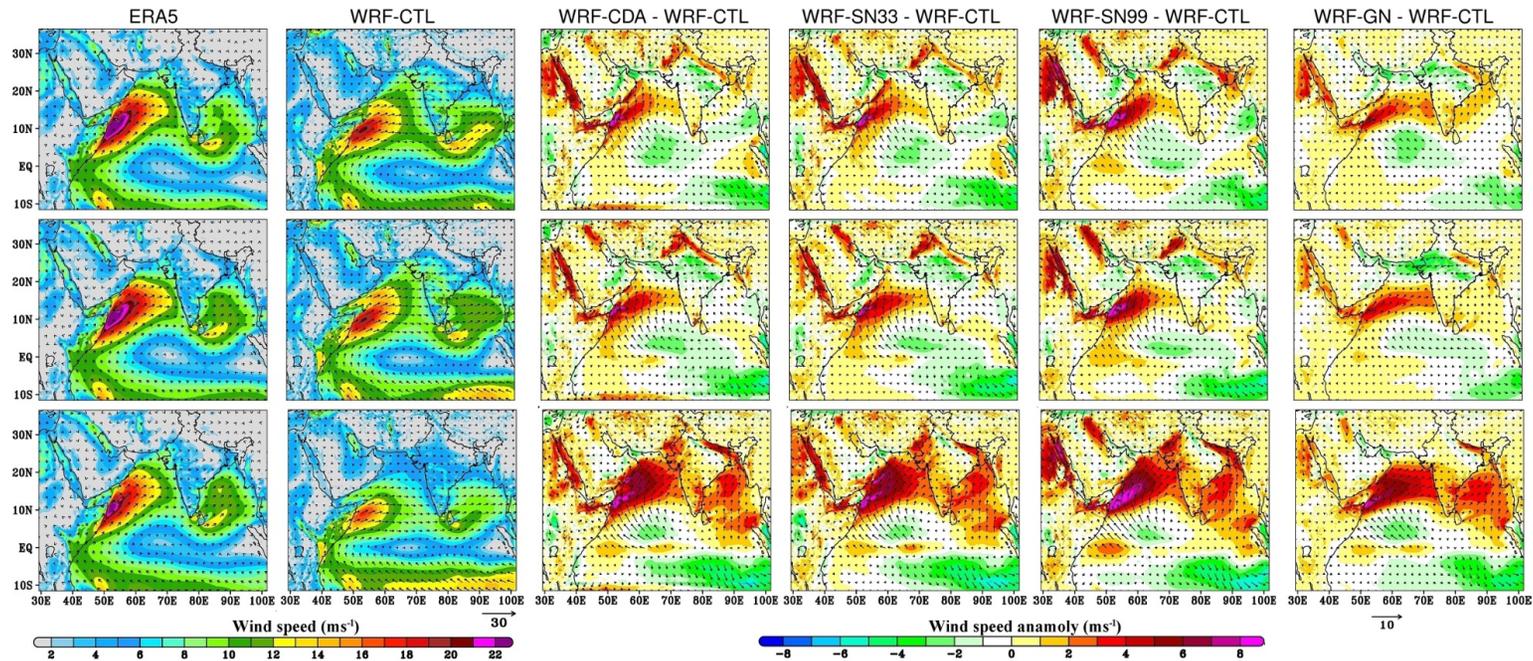

Figure 4. The mean seasonal winds (m/s) at 950 hPa of ERA5 (First column) and WRF-CTL (Second column), and the differences in the mean zonal wind shear (m/sec) between the WRF-CDA and WRF-CTL (Third column), WRF-SN33 and WRF-CTL (Fourth column), WRF-S99 and WRF-CTL (Fifth column), and WRF-GN and WRF-CTL (Sixth column) during normal (2016), excess (2013), and drought (2009) Indian Summer Monsoon seasons. First row is for normal, second row is for excess, and third row is for drought Indian Summer Monsoon seasons.

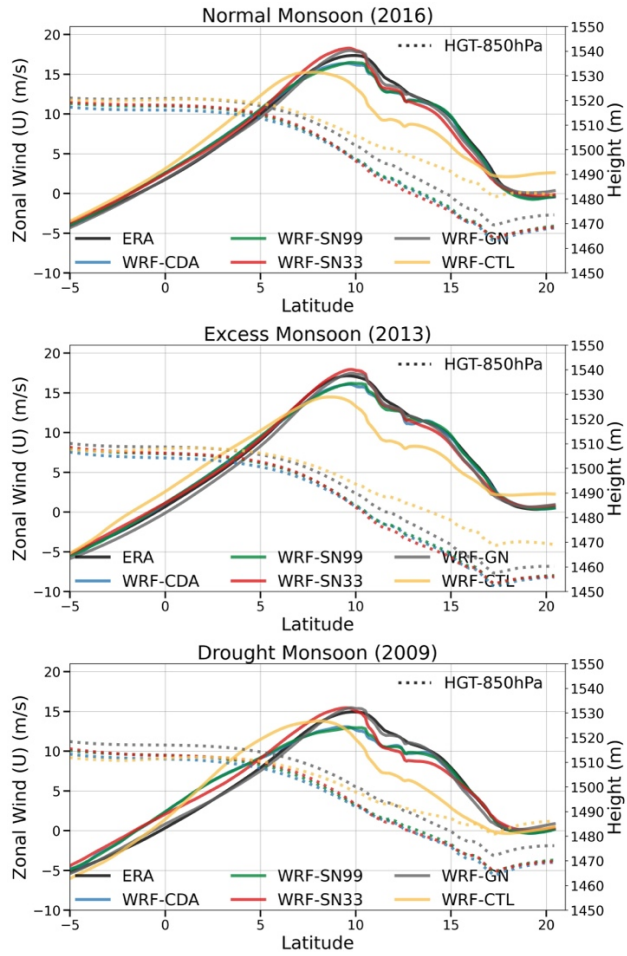

Figure 5: Latitudinal section of the mean seasonal zonal winds (m/s) and geopotential heights (m) at 850 hPa averaged over the longitudes 50E-60E during normal (2016), excess (2013), and drought (2009) Indian Summer Monsoon seasons.

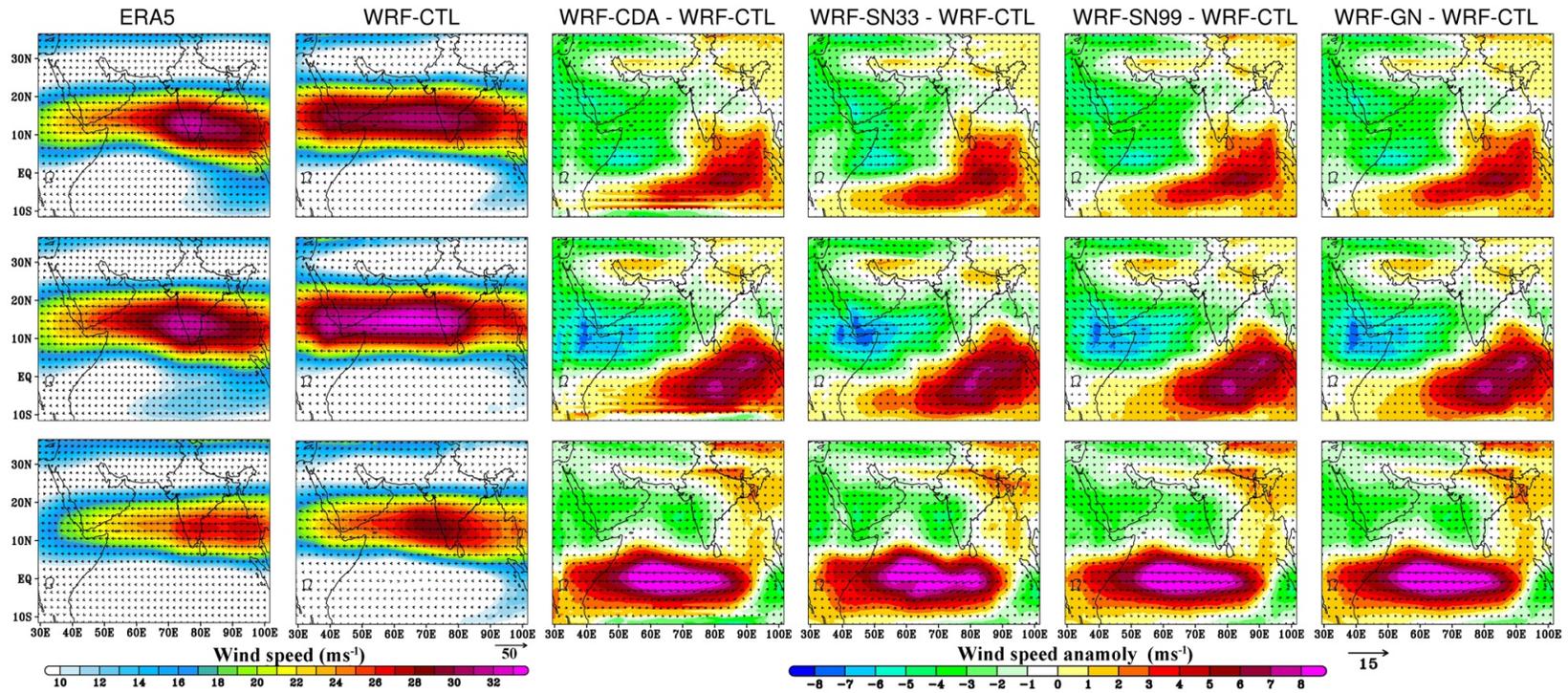

Figure 6: The mean seasonal winds (m/s) at 100 hPa of ERA5 (First column) and WRF-CTL (Second column), and the differences in the mean zonal wind shear (m/sec) between the WRF-CDA and WRF-CTL (Third column), WRF-SN33 and WRF-CTL (Fourth column), WRF-S99 and WRF-CTL (Fifth column), and WRF-GN and WRF-CTL (Sixth column) during normal (2016), excess (2013), and drought (2009) Indian Summer Monsoon seasons. First row is for normal, second row is for excess, and third row is for drought Indian Summer Monsoon seasons.

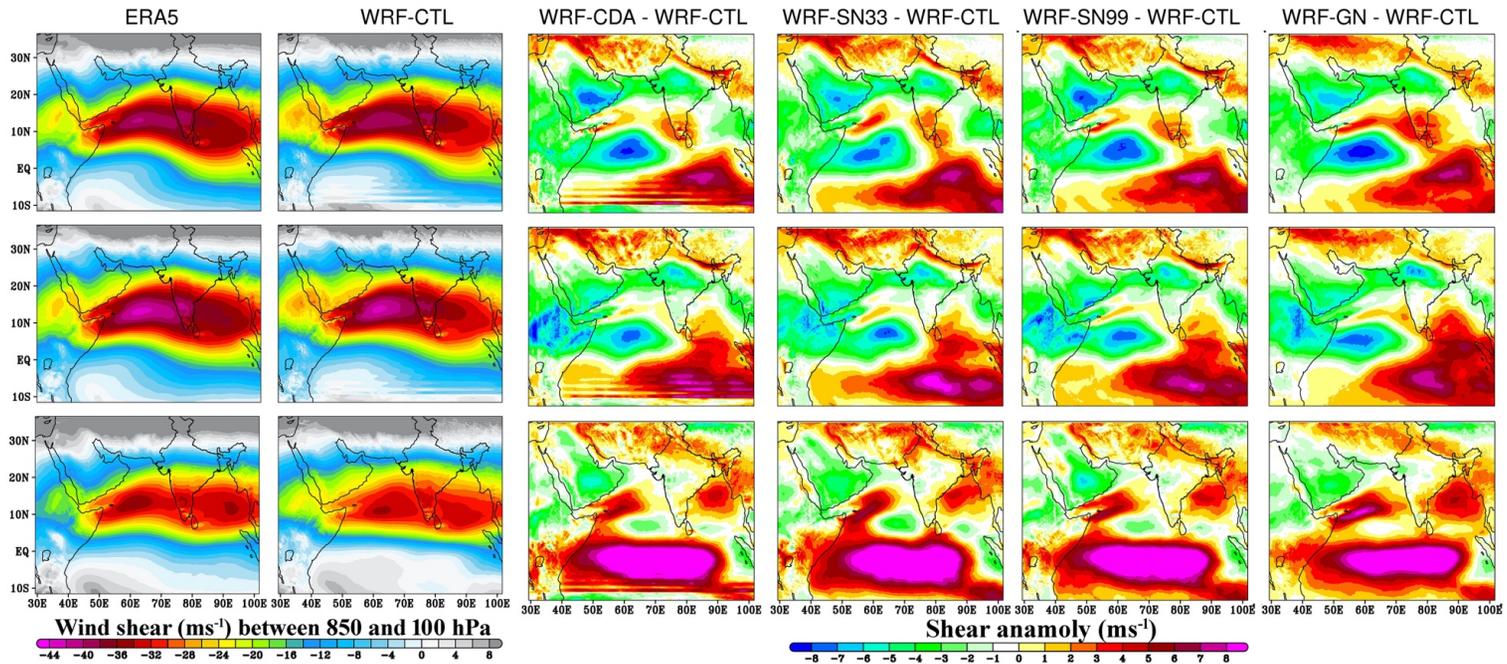

Figure 7: The mean seasonal zonal wind shear (m/s) between 850 hPa and 100 hPa levels of ERA5 (First column) and WRF-CTL (Second column), and the differences in the mean zonal wind shear (m/sec) between the WRF-CDA and WRF-CTL (Third column), WRF-SN33 and WRF-CTL (Fourth column), WRF-S99 and WRF-CTL (Fifth column), and WRF-GN and WRF-CTL (Sixth column) during normal (2016), excess (2013), and drought (2009) Indian Summer Monsoon seasons. First row is for normal, second row is for excess, and third row is for drought Indian Summer Monsoon seasons.

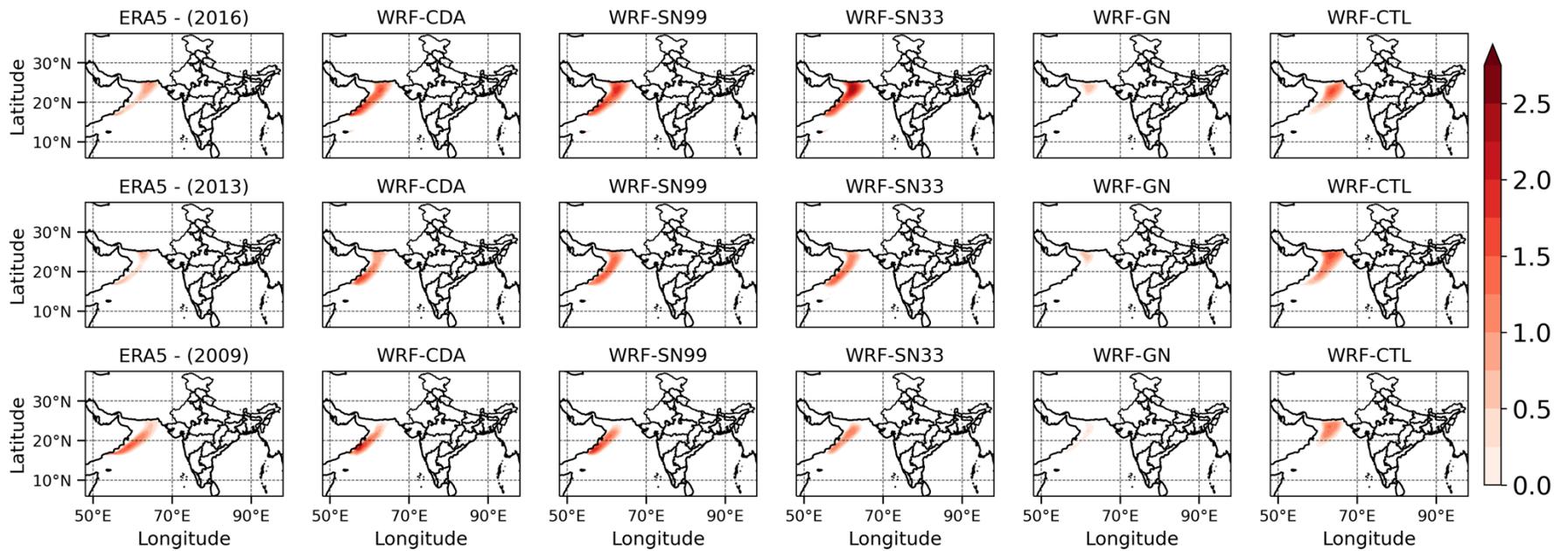

Figure 8: The mean seasonal monsoon inversion (K, differences in the temperature between the 850 and 950 [ΔT = T850 hPa – T950 hPa)]) for normal (2016), excess (2013) and drought (2009) years. First, second, third, fourth, fifth, and sixth columns are for the ERA5, WRF-CDA, WRF-SN99, WRF-S33, WRF-GN, and WRF-CTL, respectively. First row is for normal, second row is for excess, and third row is for drought Indian Summer Monsoon seasons.

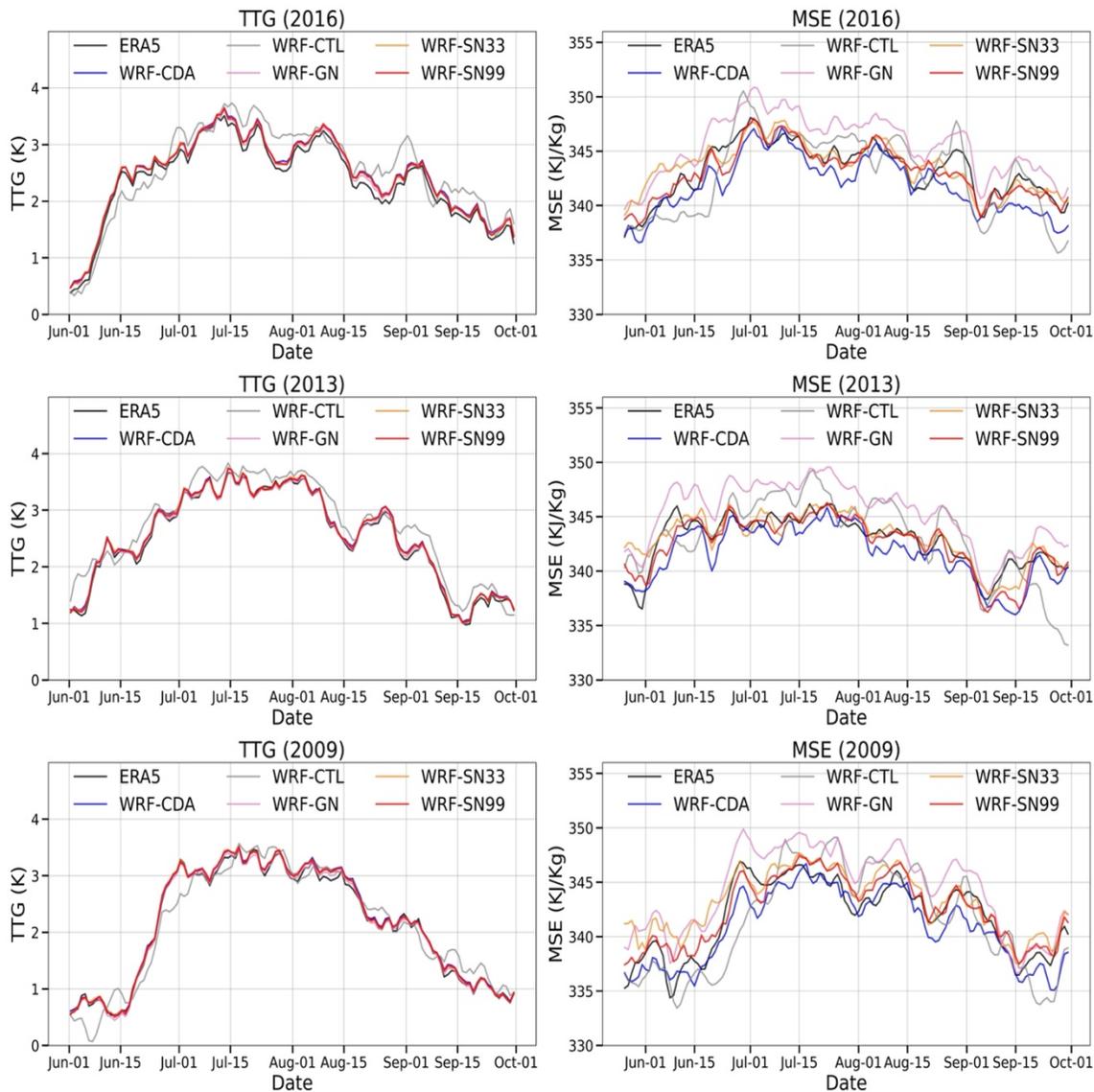

Figure 9: The time series of the tropospheric temperature (averaged between 700 hPa and 200 hPa) gradient (K) is the differences in the mean tropospheric temperature averaged over the regions 40°E–100°E; 5°N–35°N and 40°E–100°E; 15°S–5°N (left panel) for the normal (2016), excess (2013), and drought (2009) Indian summer monsoon years. The right panel is same as the left panel but for mean moist static energy (MSE, KJ/Kg) averaged over central India.

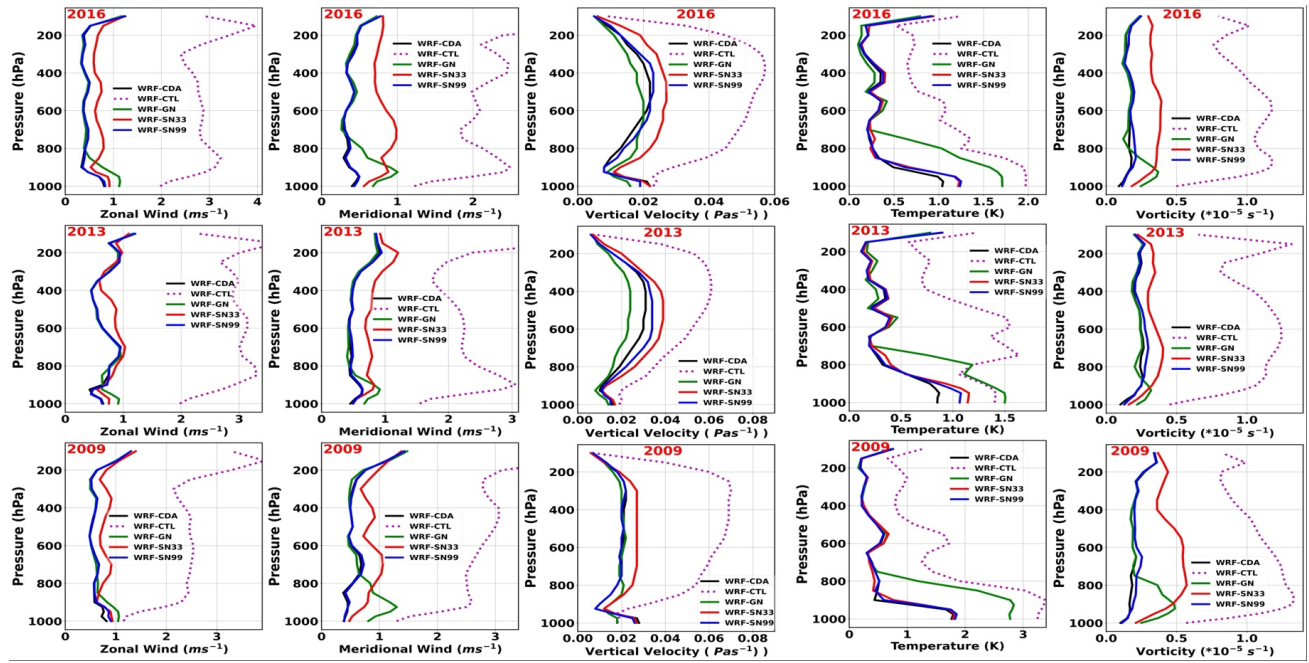

Figure 10. The root-mean-square-error computed for vertical profiles of different atmospheric variables from ERA5 and with different model simulations. The first column is for zonal wind, second column is for meridional wind, third column is for vertical wind, fourth column is for temperature, fifth column is for specific humidity, and sixth column is for relative vorticity. The first row is for normal, second row is for excess, and third row is for drought Indian Summer Monsoon seasons.

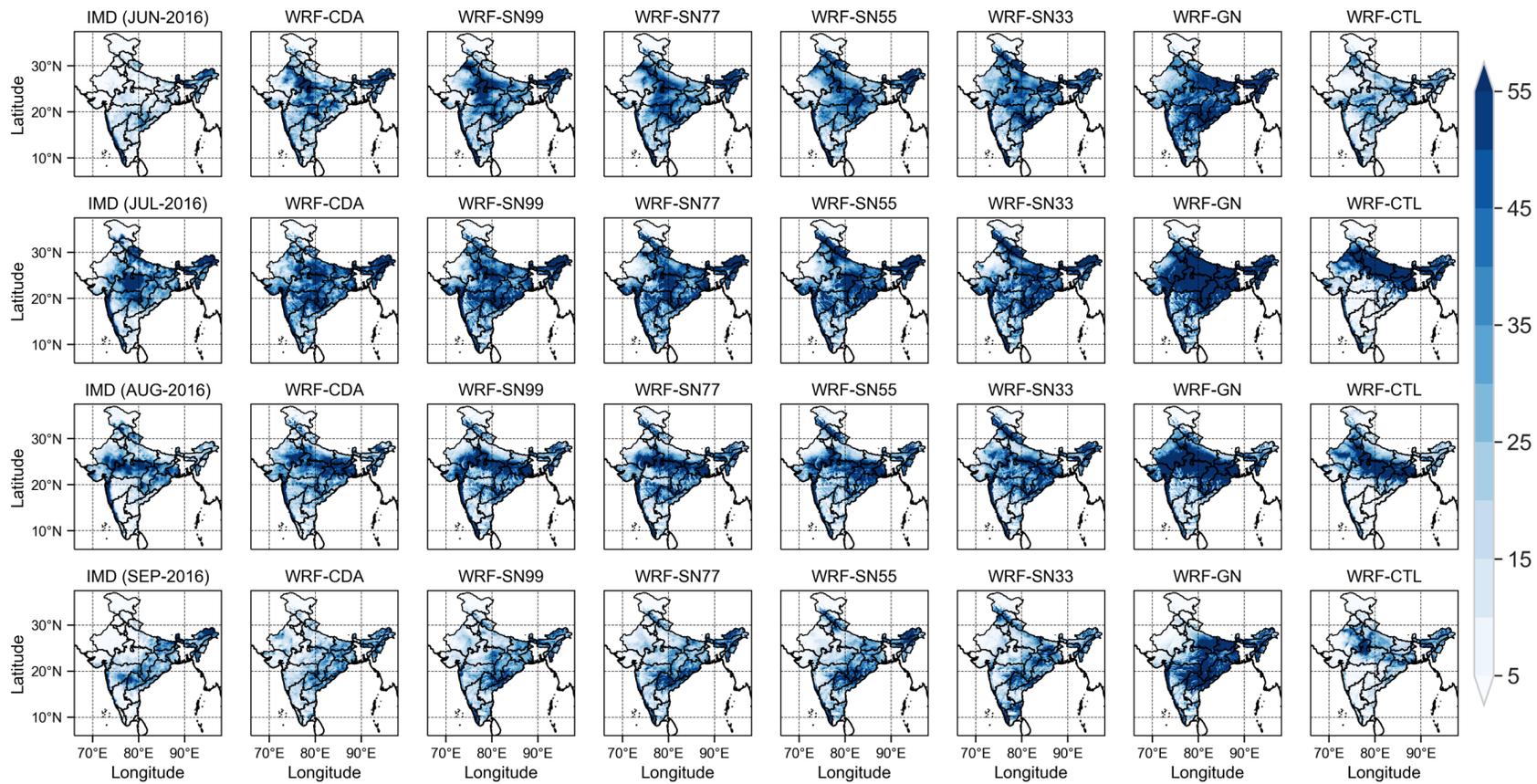

Figure S1: Spatial distribution of total monthly rainfall (cm) of normal monsoon (2016).

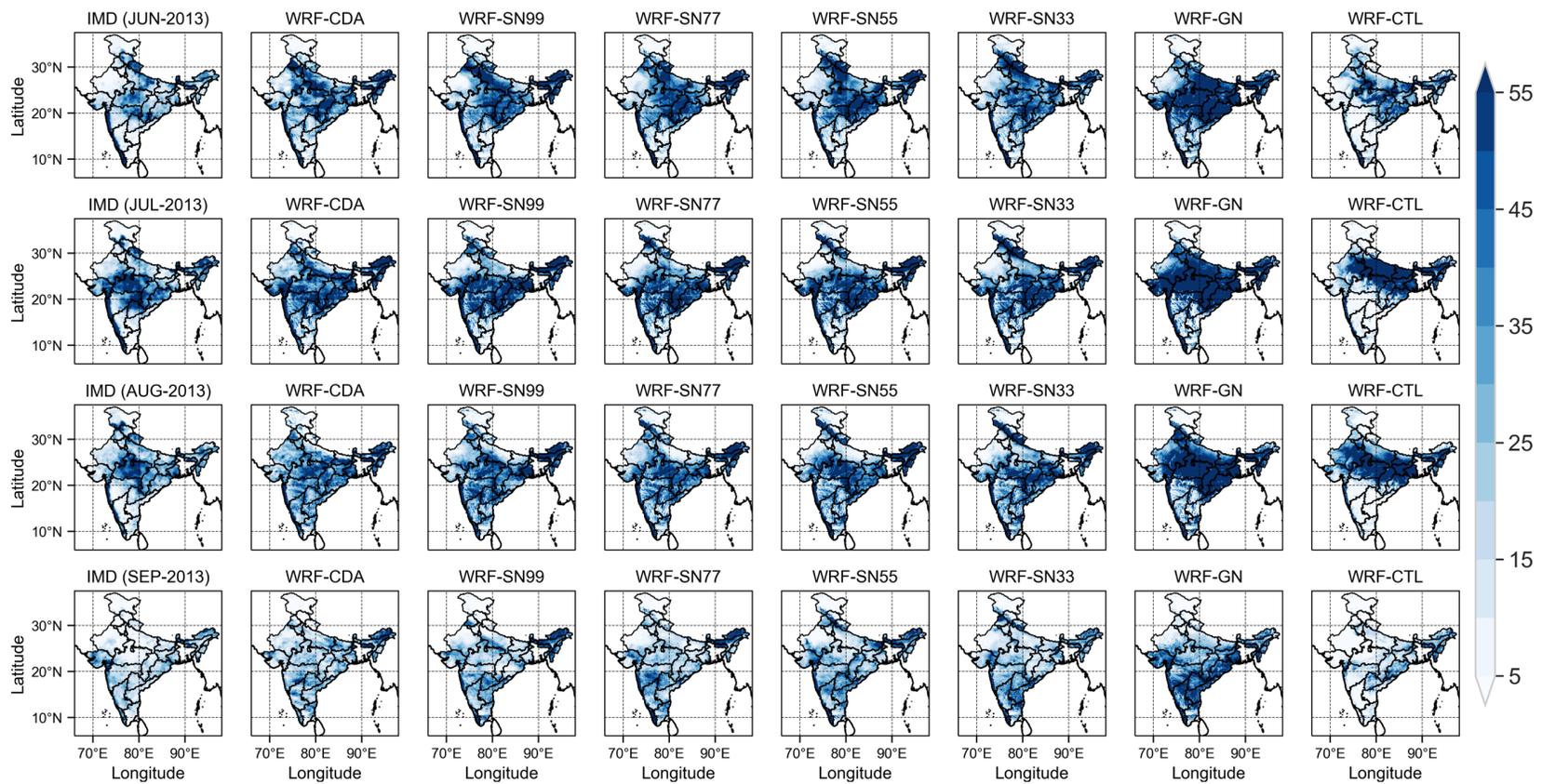

Figure S2: Spatial distribution of total monthly rainfall (cm) of excess monsoon (2013).

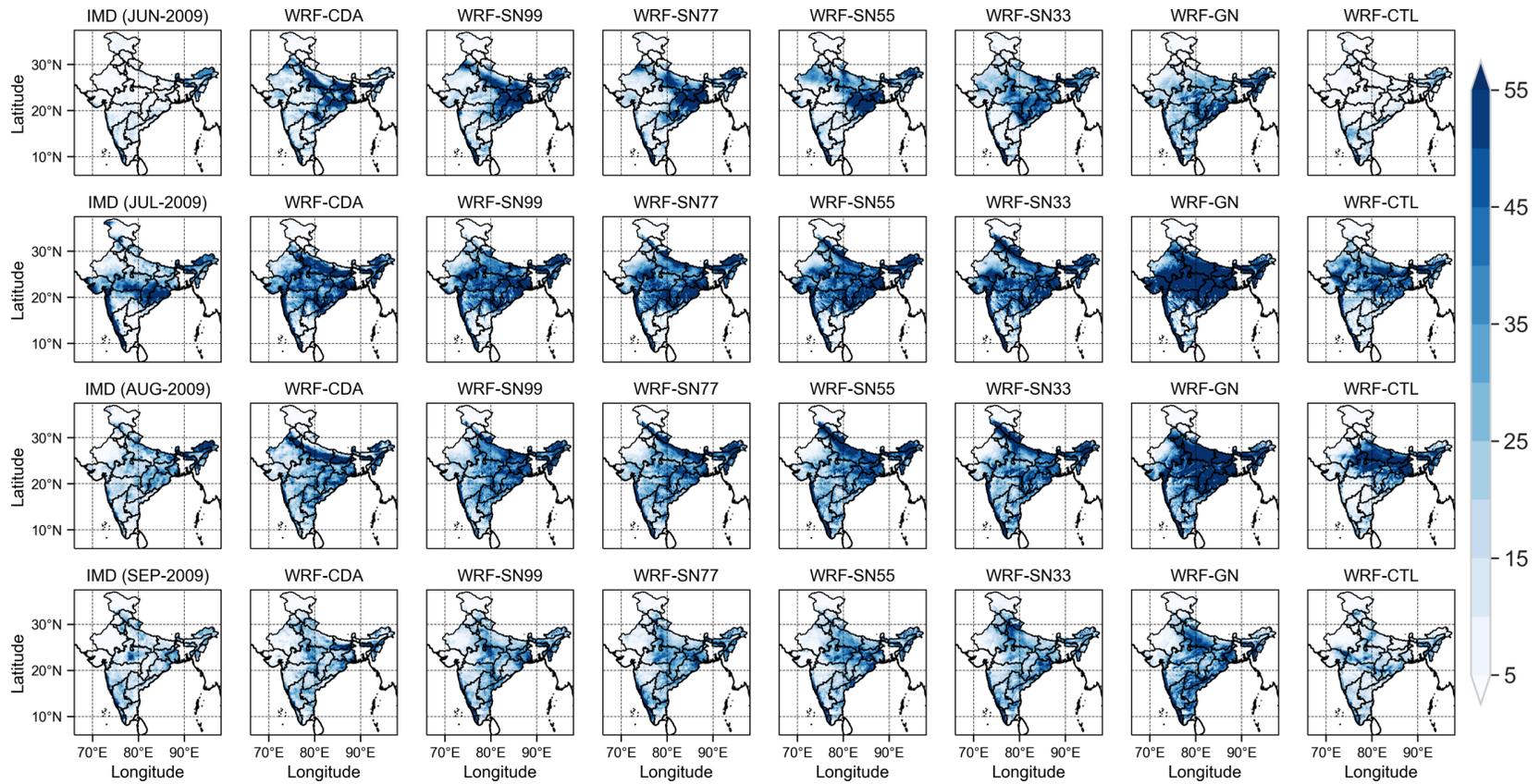

Figure S3: Spatial distribution of total monthly rainfall (cm) of drought monsoon (2009).

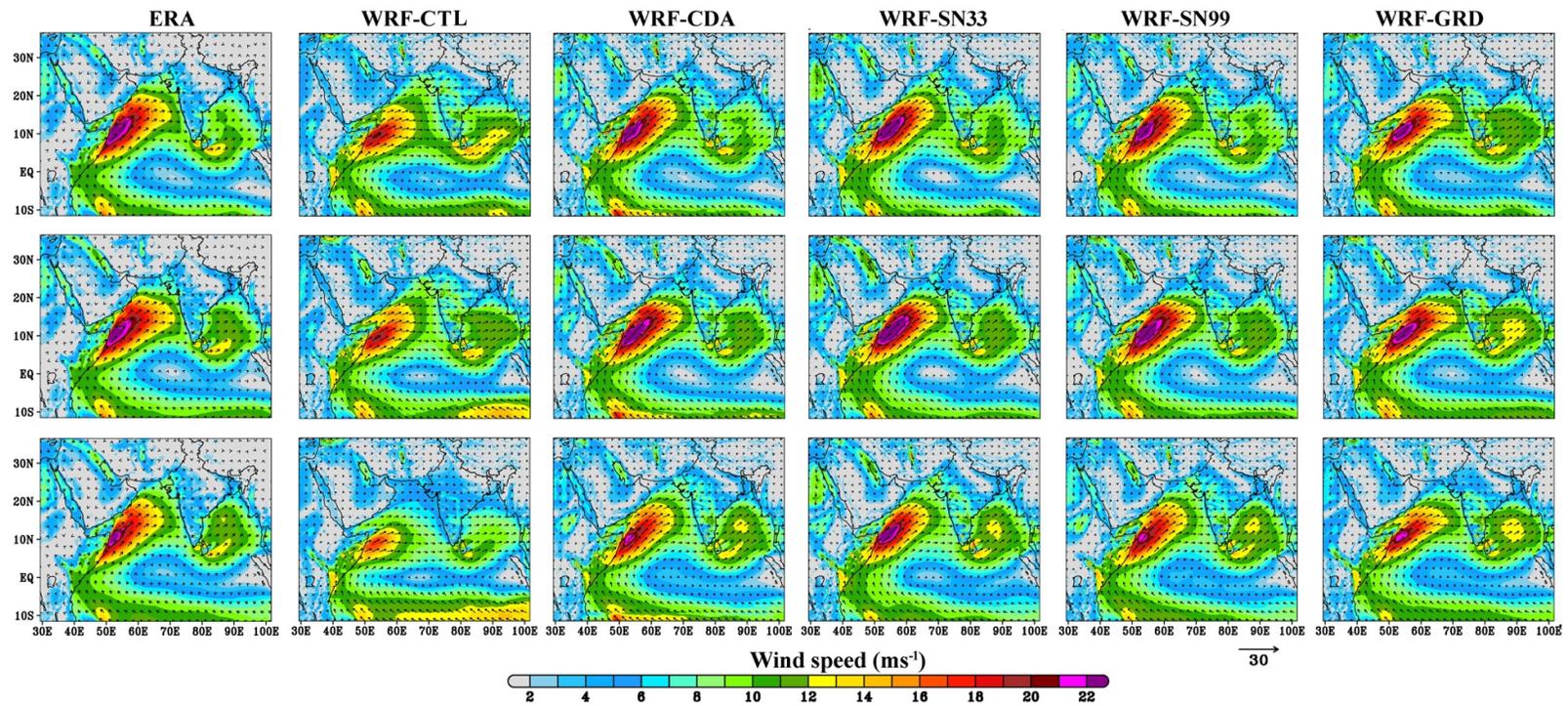

Figure S4: The mean seasonal winds (m/s) at 950 hPa during normal (2016), excess (2013), and drought (2009) Indian Summer Monsoon seasons. First, second, third, fourth, fifth, and sixth columns are for the ERA5, WRF-CTL, WRF-CDA, WRF-SN33, WRF-SN99, and WRF-GN, respectively. First row is for normal, second row is for excess, and third row is for drought Indian Summer Monsoon seasons.

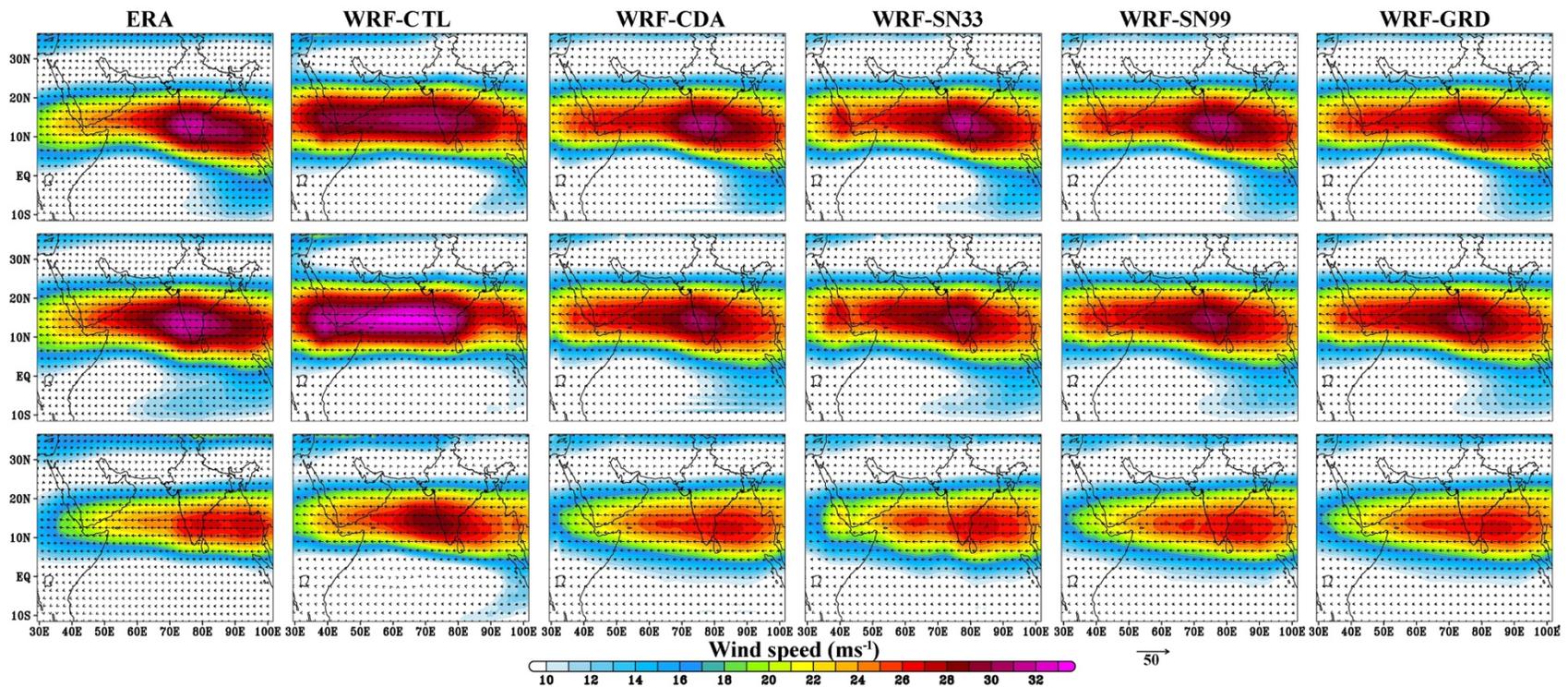

Figure S5: The mean seasonal winds (m/s) at 100 hPa during normal (2016), excess (2013), and drought (2009) Indian Summer Monsoon seasons. First, second, third, fourth, fifth, and sixth columns are for the ERA5, WRF-CTL, WRF-CDA, WRF-SN33, WRF-SN99, and WRF-GN, respectively. First row is for normal, second row is for excess, and third row is for drought Indian Summer Monsoon seasons.

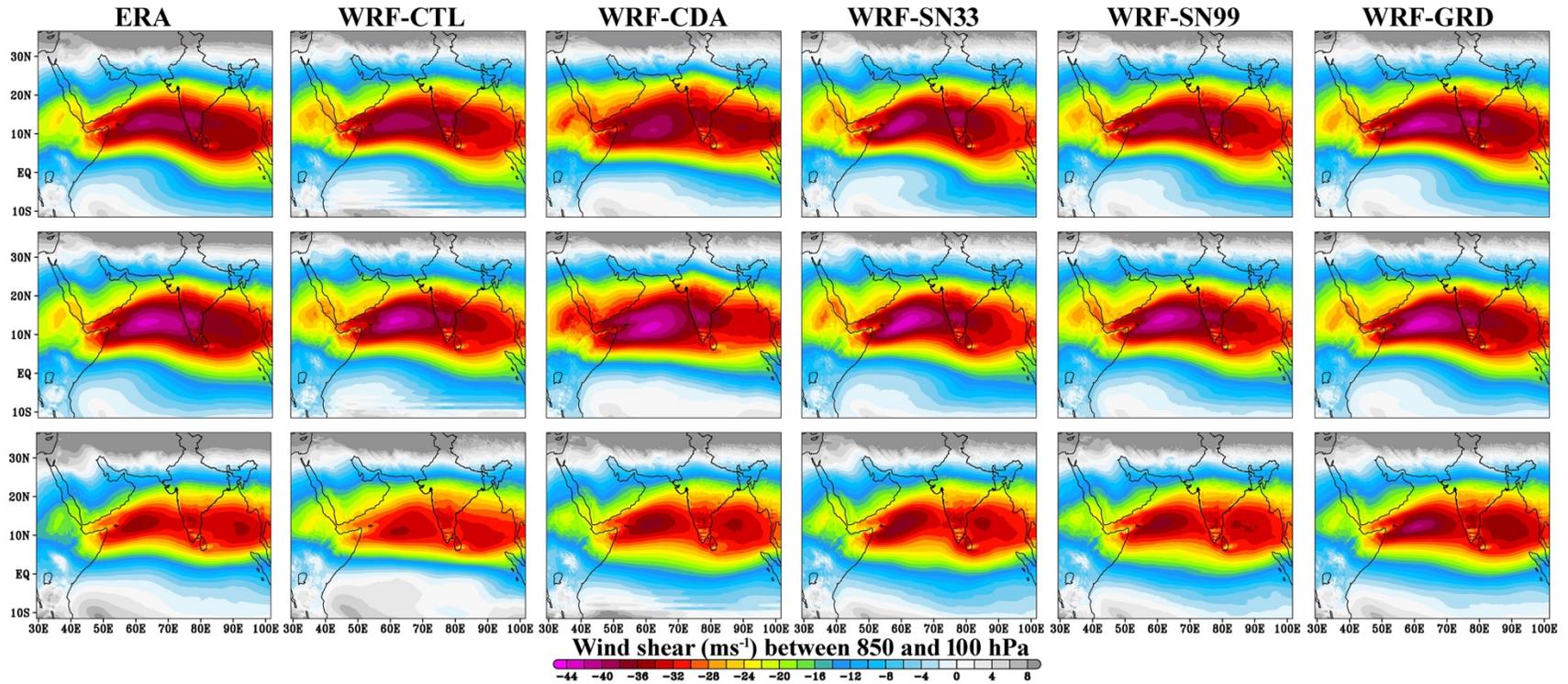

Figure S6: The mean seasonal zonal wind shear (m/s) between 850 hPa and 100 hPa levels during normal (2016), excess (2013), and drought (2009) Indian Summer Monsoon seasons. First, second, third, fourth, fifth, and sixth columns are for the ERA5, WRF-CTL, WRF-CDA, WRF-SN33, WRF-SN99, and WRF-GN, respectively. First row is for normal, second row is for excess, and third row is for drought Indian Summer Monsoon seasons.

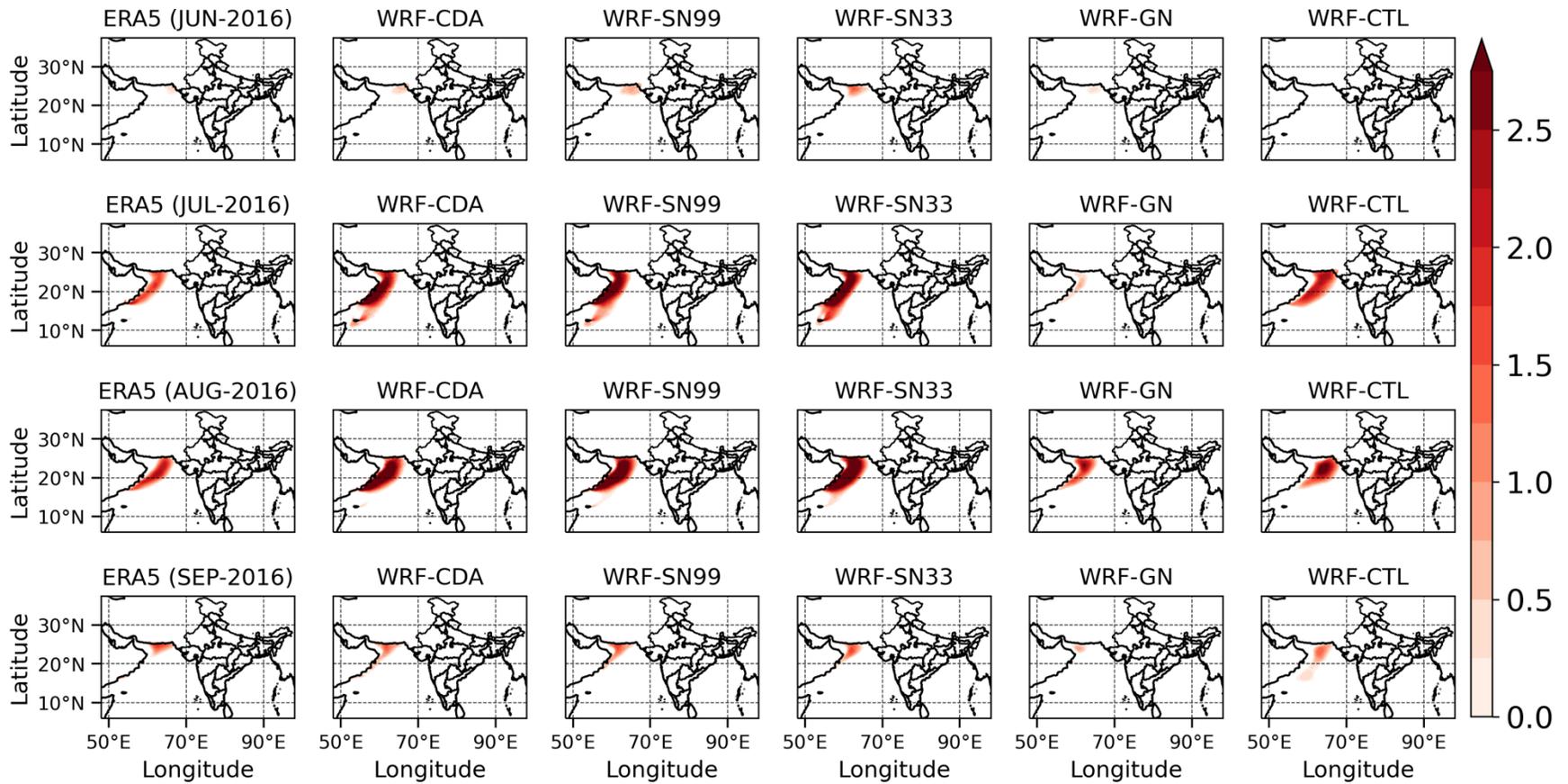

Figure S7: The mean monthly monsoon inversion (K, differences in the temperature between the 850 and 950 [ΔT = T850 hPa – T950 hPa)]) for normal (2016) Indian Summer Monsoon Season.

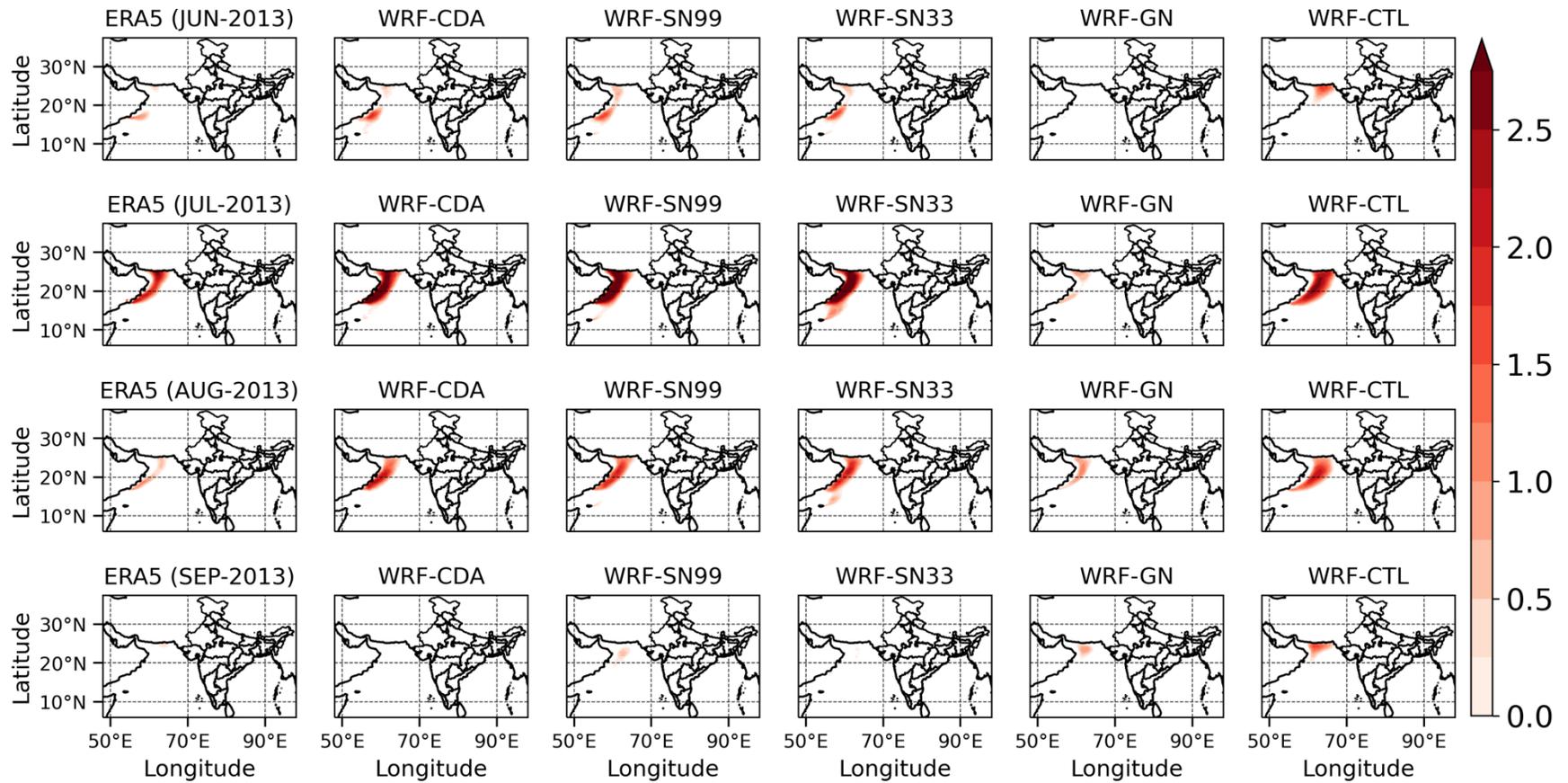

Figure S8: The mean monthly monsoon inversion (K, differences in the temperature between the 850 and 950 [ΔT = T850 hPa – T950 hPa)]) for excess (2013) Indian Summer Monsoon Season.

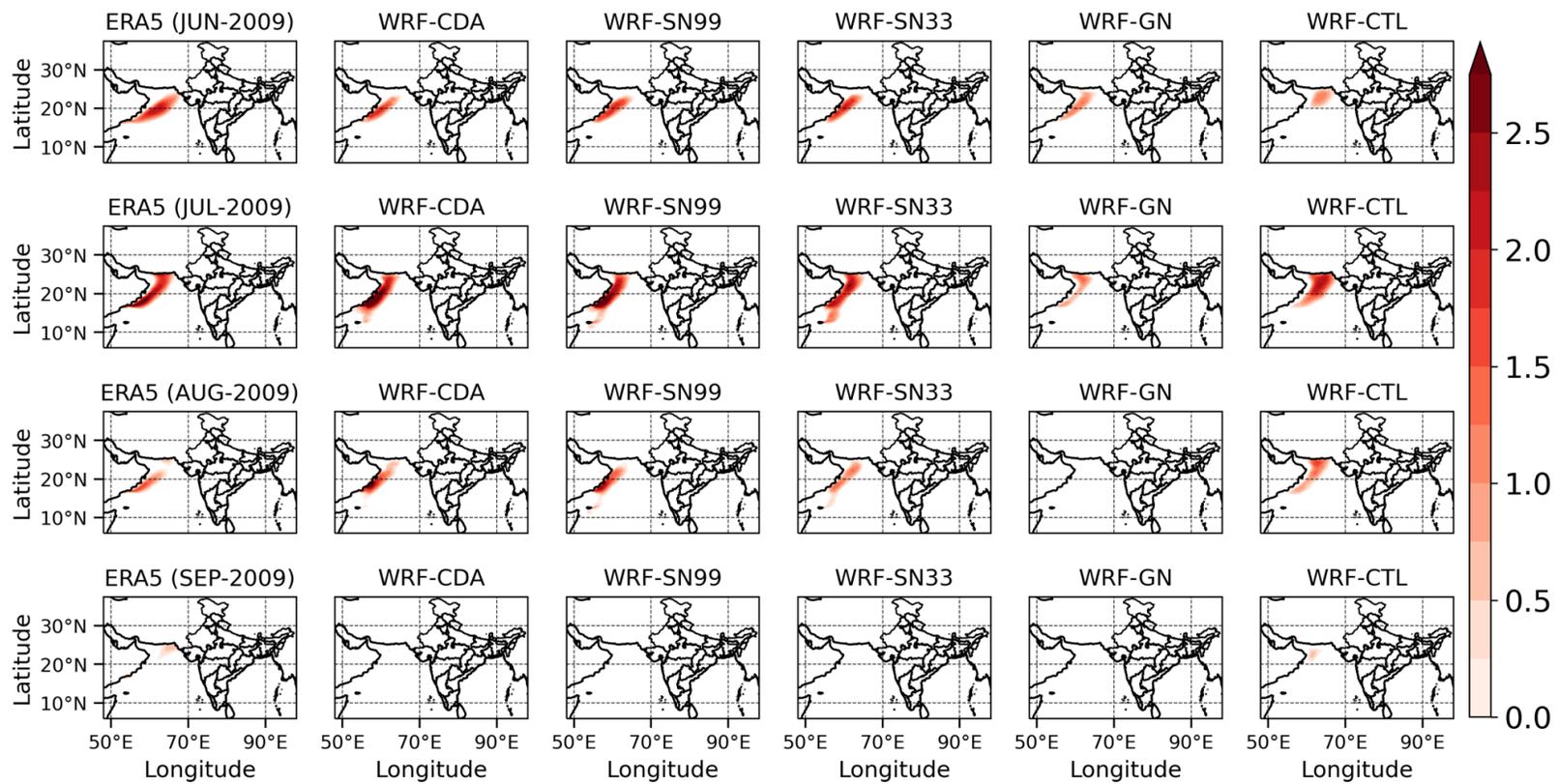

Figure S9: The mean monthly monsoon inversion (K, differences in the temperature between the 850 and 950 [ΔT = T850 hPa – T950 hPa)]) for drought (2009) Indian Summer Monsoon Season.